# High-Resolution Observations of Pickup Ion Mediated Shocks to 60 au

Bishwas L. Shrestha[1], David J. McComas[1], Eric J. Zirnstein[1], George Livadiotis[1], Heather A. Elliott[2], Pontus C. Brandt[3], Alan Stern[4], Andrew R. Poppe[5], Joel Parker[4], Elena Provornikova[3], Kelsi Singer[4], Anne Verbiscer[6], *New Horizons* Heliophysics Team

[1]Department of Astrophysical Sciences, Princeton University, Princeton, NJ 08544, USA
[2]Southwest Research Institute, San Antonio, TX 78228, USA
[3]The Johns Hopkins University Applied Physics laboratory, Laurel, MD 20723, USA
[4]Southwest Research Institute, Boulder, CO 80302, USA
[5]Space Science Laboratory, University of California at Berkley, Berkley, CA 94720, USA
[6]Department of Astronomy, University of Virginia, Charlottesville, VA 22904, USA

## Abstract

This study provides a detailed analysis of fourteen distant interplanetary shocks observed by the Solar Wind Around Pluto (SWAP) instrument onboard *New Horizons*. These shocks were observed with a pickup ion data cadence of approximately 30 minutes, covering a heliocentric distance range of ~52-60 au. All the shocks observed within this distance range are fast-forward shocks, and the shock compression ratios vary between ~1.2 and 1.9. The shock transition scales are generally narrow, and the SW density compressions are more pronounced compared to the previous study of seven shocks by McComas et al. (2022). A majority (64%) of these shocks have upstream sonic Mach numbers greater than one. In addition, all high-resolution measurements of distant interplanetary shocks analyzed here show that the shock transition scale is independent of the shock compression ratio. However, the shock transition scale is strongly anti-correlated with the shock speed in the upstream plasma frame, meaning that faster shocks generally yield sharper transitions.

---

[1] Corresponding author bishwasls@princeton.edu

# 1. Introduction

The solar wind (SW), originating from the solar corona, expands outward in all directions. Its source varies spatially and temporally and generates SW with speeds typically ranging from ~300 to 800 km s$^{-1}$ near one astronomical unit (au) from the Sun. Due to the Sun's rotation, faster parcels of SW catch up with slower ones ahead of them and create compressions. Some of these compressions develop into outward-propagating forward shocks and may even launch reverse shocks back toward the Sun in the SW frame (Richardson 2004). The interaction between fast and slow SW persists over multiple solar rotations, especially around solar minimum, forming corotating interaction regions (CIRs). These CIRs are strongest around several au from the Sun (Neugebauer 2013). Beyond ~10 au from the Sun, CIRs weaken but can merge with other CIRs and coronal mass ejections (CMEs) to form merged interaction regions (MIRs). These can be further merged into larger structures, known as globally merged interaction regions (GMIRs) (Richardson 2018). While the speed difference between fast and slow SW decreases with distance, causing the number of newly-formed shocks to decrease, speed differences continue to drive some shocks as they move through the outer heliosphere (Wang & Richardson 2002).

Interstellar pickup ions (PUIs) are formed by the ionization of interstellar neutral (ISN) atoms via charge exchange, photoionization, and electron impact ionization in the heliosphere (see the recent review by Zirnstein et al. (2022), and references therein). These ions are then picked up by the motional electric field of the SW and gyrate around the interplanetary magnetic field, creating a ring beam distribution. The newly formed PUIs gain a velocity equivalent to the relative velocity between the bulk SW and the interstellar flow. The PUI ring beam is unstable and undergoes pitch angle scattering and isotropization due to ambient and/or self-generated low-frequency electromagnetic fluctuations, forming a shell distribution. As PUIs co-move with the bulk SW, they experience a non-adiabatic cooling (McComas et al. 2021), creating a filled-shell distribution (Vasyliunas & Siscoe 1976; Chen et al. 2014; McComas et al. 2021). The filled-shell distribution contains the freshly born PUIs on the outermost layer and those formed closer to the Sun on the inner layers. The dominant source of PUIs near 1 au is interstellar helium atoms, due to their higher ionization potential (Sokół et al. 2019) and consequently smaller ionization cavity size. On the other hand, H$^+$ PUIs become the dominant internal pressure component in the distant outer heliosphere, beyond ~20 au from the Sun (Rucinski & Bzowski 1995).

The Solar Wind Around Pluto (SWAP) instrument onboard *New Horizons* measures ions in the energy range of approximately 0.021-7.8 keV/q (McComas et al. 2008). It has provided unprecedented measurement of H$^+$ PUIs now out to a heliocentric distance of around 60 au (Randol et al. 2012, 2013; Swaczyna et al. 2020; McComas et al. 2010, 2021, 2022, 2025; Shrestha et al. 2024). McComas et al. (2017) calculated moment-like H$^+$ PUI parameters by fitting a classic Vasyliunas & Siscoe (1976) model to the observed SWAP count rates from ~22-38 au, showing that H$^+$ PUIs dominate the internal pressure in the outer heliosphere by these distances. Although their model fits were generally adequate, the best-fit values of some of the fitting parameters, e.g., local ionization rate and ionization cavity size were often unphysically large or small. Swaczyna et al. (2020) reanalyzed the H$^+$ PUI observations from McComas et al. (2017) by incorporating a non-adiabatic cooling index for H$^+$ PUIs using a generalized model from Chen et al. (2014) and derived the interstellar neutral hydrogen density. They found that the interstellar neutral hydrogen density near the heliosphere is ~40% larger than the previously accepted value (Bzowski et al. 2009). McComas et al. (2021) extended the PUI observations to about 46.6 au from the Sun with improved analysis, incorporating the non-adiabatic cooling index (Chen et al. 2014; Swaczyna et al. 2020) as a free parameter. They discovered that the daily averaged PUI distribution beyond ~22 au shows additional heating of PUIs besides just adiabatic cooling. Additionally, they used a superposed epoch analysis of the low-resolution (1 day) SWAP data available at those distances to examine the "average behavior" of 39 interplanetary shocks over that interval. Livadiotis et al. (2024) examined the PUI cooling index and showed that it is related to the polytropic index of PUIs that describes their thermodynamic polytropic processes. They further developed the connection between the cooling index and thermodynamic kappa parameters. Most recently, (McComas et al. 2025) provided the radial profile of PUIs in the outer heliosphere from ~22-60 au by stitching together older daily averaged data and newer ~30-minute resolution data. They provided the radial gradients of all PUI parameters and the ratios to their SW counterparts.

The first in-depth study of distant interplanetary shock using SWAP data was conducted by Zirnstein et al. (2018), where they analyzed a strong shock at around 34 au from the Sun. They discovered that PUIs are preferentially heated across the shock compared to SW protons. Their study also revealed the formation of high-energy H$^+$ PUI tail downstream of the shock, which accounted for about 20% of the total downstream energy flux. McComas et al. (2022) provided a detailed analysis of 7 distant interplanetary shock waves (6 fast-forward and one fast-reverse) using the then available high-resolution SWAP data (~30 min resolution) from ~49.5 – 52 au. Their study was the first to resolve the shock transitions at distant interplanetary shocks and confirmed the preferential heating of PUIs across these shocks. Shrestha et al. (2024) examined 5 distant interplanetary shocks with distinct suprathermal $H^+$ PUI tails in the downstream distribution observed over 24-37 au. They found that the suprathermal tail is observable with shock compression ratios from 1.4 to 3.2, and the number density of the $H^+$ PUI tail increases with stronger shocks. Their findings also showed that the number density fraction of the PUI tail aligns well with the theory of particle reflection from the electrostatic cross-shock potential (CSP).

PUIs are likely preferentially heated at the heliospheric termination shock (HTS) (Zank et al. 1996; Lee et al. 1996; Richardson et al. 2008a; Mostafavi et al. 2018; Zirnstein et al. 2021) and are believed to be the primary source of ~0.5-7 keV ions in the heliosheath (Zank et al. 2010; Shrestha et al. 2020, 2021, 2023; Zirnstein et al. 2021; Kornbleuth et al. 2023; Gkioulidou et al. 2022; Baliukin et al. 2022). Consequently, PUIs play a crucial role in the pressure balance in the heliosheath (Decker et al. 2008; Dialynas et al. 2020) . They are the parent population of energetic neutral atoms (ENAs) created beyond the HTS, being observed by NASA's Interstellar Boundary Explorer (IBEX) (McComas et al. 2009) and soon to be observed by NASA's Interstellar Mapping and Acceleration Probe (IMAP) (McComas et al. 2018).

This study presents a detailed analysis of the fourteen high-resolution shocks observed by the SWAP instrument on *New Horizons* in the distant outer heliosphere. These shocks were observed from 24 December 2021 to 13 September 2024 over a heliocentric distance of approximately 52 to 60 au. These shocks are relatively weak, with the compression ratios ranging from ~1.2 to 1.9. This study also incorporates the six high-resolution fast-forward shocks (the reverse shock S5 is excluded), previously analyzed by McComas et al. (2022) to derive the statistical relation between different shock parameters.

## 2. Data and Methodology

This study uses the *New Horizons* SWAP high-resolution data to examine the properties of the PUI-mediated shocks in the distant outer heliosphere. This dataset includes the high-resolution observations of PUIs from 19 February 2021 to 24 December 2021 (~49.5-52 au) analyzed by McComas et al. (2022) and the newer observations that extend from 24 December 2021 to 13 September 2024 (~52-60 au). We use the shock list from McComas et al. (2025) that identified the shocks in the newer high-resolution SWAP data if there is (1) a sharp jump in the SW speed within a time window of ~30 minutes and (2) a visibly distinct jump in the PUI density corresponding to the SW speed jump. Identifying the shocks without magnetic field measurements is challenging, and our goal is not to examine ideal shocks from an MHD perspective. Instead, we characterize shocks in terms of changes in SW speed and PUI density to understand how PUI mediates these structures.

Below we define how the shock compression ratio and the shock speed are calculated:

I. Shock compression ratio ($r_{\text{comp}}$): We define the shock compression ratio as the ratio of PUI density between downstream ($n_2$) and upstream ($n_1$) of the shock, given by

$$r_{\text{comp}} = \frac{n_2}{n_1}. \quad (1)$$

We use the PUI density compression rather than the SW density (or total density) to define the shock compression ratio because the SW density does not appear to show a distinct compression downstream of distant interplanetary shocks as exhibited by PUIs. Rather, they exhibit small-scale fluctuations throughout the shock transition (McComas et al. 2022; Shrestha et al. 2024), which may not be associated with shocks.

II. Shock speed ($V_{\text{sh}}$): The shock speed is calculated by using the PUI density compression from upstream to downstream and assuming that the upstream and downstream plasma flow ($u_1$ and $u_2$) are parallel to the shock normal. The shock speed in the solar inertial frame is given by

$$V_{\text{sh}} = \frac{n_2 u_2 - n_1 u_1}{n_2 - n_1}, \quad (2)$$

where $u_1$ and $u_2$ are upstream and downstream SW speeds in the solar inertial frame, respectively. Note that the SWAP data processing assumes the SW ions and PUIs are co-moving, and the SW speed provided in the SWAP data release is in the solar inertial frame (not in the spacecraft frame).

### 3. Results and Discussions

Figure 1 shows the overview of the SWAP high-resolution observations from 19 Feb 2021 to 13 September 2024, which covers the heliocentric distance range of ~49.5-59.9 au. The top four panels show the SW speed, density, temperature, and thermal pressure, while the bottom four show the PUI density, temperature, thermal pressure, and cooling index (or polytropic index). The fourteen new shocks (numbered 8-21), identified by using the criteria described in Section 2, along with the seven shocks (numbered 1-7) analyzed in McComas et al. (2022), are marked by gray vertical lines and sequential numbers. The observation timing, the radial distances of these shocks, and the shock parameters are listed in Table 1 (see Section 3.1 for details).

Notably, the time interval between the observation of shocks 11, 12, 13, and 14 are ~26, ~24, and ~23 days, respectively. Furthermore, the intervals between shocks 18, 19, and 20 are ~30 and ~27 days, respectively. These time intervals are comparable to the average Carrington Rotation period of ~27.3 days, indicating that these shocks likely originate from CIRs. A detailed study of the connection between these shocks and CIRs observed at 1 au will be the subject of future research.

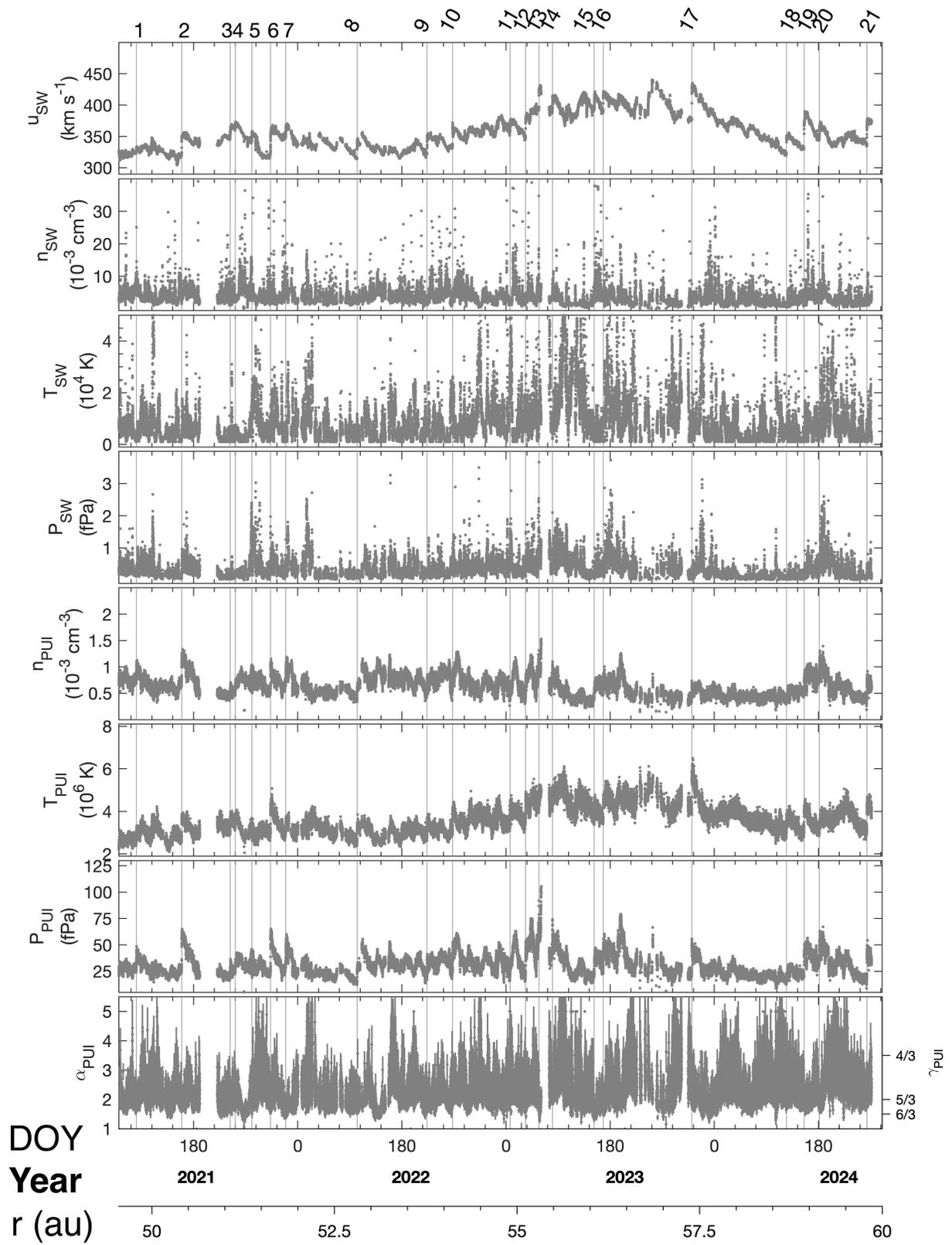

*Figure 1. High-resolution SWAP data from 19 February 2021 to 13 September 2024 (~49.5-59.9 au). The top four panels represent the SW properties (bulk speed, density, temperature, and thermal pressure), and the bottom four represent the PUI quantities (density, temperature, thermal pressure, and cooling index (or polytropic index)). The solid vertical gray line marks the position of the shocks. The numbers 1-7 represent the shock analyzed by McComas et al. (2022), and 8-21 are shocks analyzed in detail in this paper. Figure adapted from McComas et al. (2025).*

### 3.1 Variation of PUI Properties Across Distant Interplanetary Shock

We now examine SW and PUI properties across each of the 14 new high-resolution shocks in detail. Figure 2 shows an overview of shock S8 observed by SWAP on 2022 April 07 (DOY 97) at 15:52:35 UTC at a radial distance of 52.83 au. The solid gray vertical line represents the position of the shock, and the shaded gray regions on each side represent the upstream and downstream intervals of ~6 hr, where the SW speed and PUI density are relatively stable and which we use to take averages of the upstream and downstream quantities. The SW speed has an ~8.2% jump from 316 - 342 km s$^{-1}$, taking a few hours to pass over the spacecraft, which travels radially outward at a speed of ~14 km s$^{-1}$. The jump in the SW speed is followed by a slight dip and then a small subsequent rise, after which the speed remains relatively stable. The SW temperature starts to increase before the shock arrival, decreases afterward, and shows small-scale fluctuations throughout the 20-day interval. The H$^+$ PUIs are compressed and heated downstream of the shock, producing a significantly enhanced PUI pressure downstream. This enhancement is relatively constant with small point-to-point variations and lasts about five days after the shock. No jump in the SW density occurs from upstream to downstream; instead, a jump occurs around half a day after the shock. Like the SW temperature, the SW density also shows many small-scale fluctuations over the entire 20-day interval, likely not associated with the shock itself. Similar behavior of H$^+$ SW ions around shocks in the distant outer heliosphere was reported in previous Voyager (Richardson et al. 2008, Lazarous et al. 1999) and SWAP observations (McComas et al. 2022, Shrestha et al. 2024).

This strange behavior of SW density across the distant interplanetary shock is not well understood. As pointed out by previous authors (McComas et al. 2022; Shrestha et al. 2024), the reason for this could be due to (i) differential flow between SW ions and PUIs that occur in the upstream shock foot, (ii) non-shock-related smaller structures affecting the SW ions, making them less responsive to the shock in idealized scenarios, and possibly other reasons. It is plausible that the reason for this can be narrowed down with the use of sophisticated numerical simulations, but this is beyond the scope of this paper. Interestingly, the PUI cooling index ($\alpha_{PUI}$) remains relatively stable before and after the shock, with an average value of around 2. The compression ratio of this shock based on the PUI density compression is 1.50, and the shock speed in the solar inertial frame is 393 km s$^{-1}$. The upstream and downstream H$^+$ SW and PUI quantities are listed in Table 1.

Figure 3 shows an overview of shock S9, which exhibits many of the same features of shock S8. These include clear enhancements in PUI density, temperature, and pressure downstream, but a relatively stable cooling index throughout the shock transition. A noticeable feature of this shock is that the PUI density profile after the shock shows many quasi-periodic structures with a period of ~2 days. Also, the SW density shows a jump downstream but with a much wider transition than the SW speed profile. The SW density also exhibits a large-scale structure around four days after the shock. The compression ratio and the shock speed in the solar inertial frame are 1.45 and 403 km s$^{-1}$, respectively.

Figure 4 shows an overview of shock S10, which is similar to shock S8; however, it exhibits some different features. The SW speed shows a very sharp jump after the shock, followed by many fluctuations for almost a day-long period. A similar fluctuation in the PUI density correlated with the SW speed is also evident. In addition, an enhancement in the PUI density is observed around three days after the shock, which lasted around 9 days and seems anticorrelated with the PUI temperature. The compression ratio and the shock speed in the solar inertial frame are 1.51 and 400 km s$^{-1}$, respectively.

Figure 5 shows an overview of shock S11. This shock is relatively wide compared to the previous three shocks and the SW speed transition spans ~6 hr. Surprisingly, the SW density is relatively stable from upstream to downstream and then shows a large enhancement around 3 days after the shock. On the other hand, the SW temperature shows a large enhancement around a day before the shock. The PUI density, temperature, and pressure show a small jump from upstream to downstream. The PUI density jump is followed by a large-scale enhancement around 4 days after the shock. Interestingly, the PUI cooling index increases right after the shock and returns to the upstream value around 4 days after the shock. The compression ratio and the shock speed in the solar inertial frame are 1.16 and 463 km s$^{-1}$, respectively. Of all those shocks studied here, this is the shock with the weakest compression in the PUI density in SWAP's high-resolution observations.

Figure 6 shows an overview of shock S12, which exhibits a relatively steep jump in the SW speed from 349 to 378 km s$^{-1}$. Interestingly, the SW density shows a clear jump occurring with the speed jump. The SW density jump is followed by a large-scale enhancement (with many small-scale variations) around 3 days after the shock, which lasted around 9 days. The PUI density shows a relatively wider transition from upstream to downstream compared to the SW density. In addition, the PUI density exhibits a gradual increase up to around 10 days after the shock. Unlike earlier shocks, the PUI cooling index decreased around half a day after the shock. The compression ratio and the shock speed in the solar inertial frame are 1.74 and 417 km s$^{-1}$, respectively. Furthermore, the density compression of SW is 2.5, which is ~47% larger than the PUI density compression. We interpret this not as a direct measure of the shock compression ratio, but rather due to another dynamic structure that influenced the SW compression in addition to the shock.

Figure 7 shows an overview of shock S13, where the SW speed jumps quickly from ~396 km s$^{-1}$ before the shock to ~424 km s$^{-1}$ after the shock. The SW density shows a behavior similar to a reverse shock: a decrease in the density from before to after the shock (see Figure 8 in McComas et al. (2022)). However, the PUI density shows a broad jump around a day before the shock, followed by a quick jump around the shock (spanning only a few data points). The quick jump in the PUI density near the shock makes it very difficult to select the upstream and downstream regions for doing a Rankine-Hugoniot analysis to derive the shock parameters. For this reason, this shock is excluded from further analysis presented in the sections below. The PUI temperature still exhibits a distinct jump from upstream to downstream.

Figure 8 shows the overview of shock S14, which is quite different from the previous shocks discussed above. The SW speed shows a sharp jump followed by a gradual decrease for around half a day, then a very slow increase afterward for around a day. The SW density also shows a sharp jump around an hour before the shock, followed by an even larger jump around half a day after. The SW temperature and thermal pressure are also enhanced after the shock. The PUI density exhibits a similar profile as the SW speed around the shock, while the PUI temperature is relatively stable for around half a day after the shock. The PUI cooling index is relatively stable around the shock and shows an increase only around 10 days after the shock. This behavior of the cooling index is similar to the finding of McComas et al. (2022) in an earlier analysis of the first seven high-resolution shocks. The compression ratio and the shock speed in the solar inertial frame are 1.21 and 511 km s$^{-1}$, respectively.

Figure 9 shows an overview of shock S15, which exhibits a relatively broad transition around the shock in the SW speed profile. The SW density does not show a clear density compression from upstream to downstream but shows two large-scale structures: one starts around 2 days before the shock and the next one starts around 5 days after the shock. Heating of the SW from upstream to downstream is also not evident. The PUI density, temperature, and pressure show a clear jump from upstream to downstream with a much wider transition, similar to the SW speed. The PUI cooling index decreases around 2 days before the shock and remains relatively stable after the shock. The compression ratio and the shock speed in the solar inertial frame are 1.35 and 478 km s$^{-1}$, respectively.

Figure 10 shows an overview of shock S16, which has a very sharp jump in the SW speed (around 1 hr transition). The SW density shows a small jump downstream, followed by many large structures afterward. A jump in the SW temperature from upstream to downstream is not evident; however, the temperature also exhibits many large structures further downstream. The large structures in temperature appear to be anti-correlated with the SW density structures, suggesting some sort of pressure balance in the SW component. The PUI density, temperature, and thermal pressure show clear jumps from upstream to downstream with relatively stable values around half a day before and after the shock. The PUI parameters also show many quasi-periodic structures further downstream. The compression ratio and the shock speed in the solar inertial frame are 1.40 and 483 km s$^{-1}$, respectively.

Figure 11 shows an overview of shock S17, similar to shock S9 in many aspects. The SW density does not show a jump from upstream to downstream but exhibits many small-scale structures throughout. The PUI cooling index gradually declines about 5 days before the shock, experiences a steeper drop during the shock transition, and then almost stabilizes afterward. The compression ratio and the shock speed in the solar inertial frame are 1.51 and 528 km s$^{-1}$, respectively. This is the fastest shock observed so far in the SWAP's high-resolution data.

Figure 12 shows an overview of shock S18, which is somewhat similar to shock S17. The SW density and temperature remain relatively flat throughout the shock transition. Three distinct large-scale structures in the PUI density and thermal pressure are visible after the shock, with progressively increasing width. The compression ratio and the shock speed in the solar inertial frame are 1.23 and 466 km s$^{-1}$, respectively.

Figure 13 shows the overview of shock S19, which is, again, similar to shocks S17 and S18. The SW speed shows many small jumps until around two days after the shock, after which it is relatively stable for a longer period. The three large structures in PUI density and temperature are still distinct but show small fluctuations within the structures. The PUI cooling index shows a small decrease from upstream to downstream and remains relatively uniform afterward. The compression ratio and the shock speed in the solar inertial frame are 1.54 and 448 km s$^{-1}$, respectively.

Figure 14 shows the overview of shock S20, which has a wider transition in the SW speed. The SW density shows many small-scale fluctuations throughout but without a clear compression downstream. The jump in the PUI density and temperature is relatively broad and correlates with the SW speed jump. The large-scale structures in the PUI density and temperature are not distinctive and appear to exhibit many small-scale fluctuations after the shock. The PUI cooling index slowly decreases until around 1 day after the shock, after which it starts to increase slowly, followed by a faster increase around 10 days after the shock. The compression ratio and the shock speed in the solar inertial frame are 1.34 and 429 km s$^{-1}$, respectively.

Figure 15 shows the overview of shock S21, which is the last shock analyzed in this paper. The SW speed has a similar profile after the shock transition to shock S19. The SW density appears to increase around a half day before the shock, followed by a distinct jump around the timing of the speed jump. The jump in the PUI density is broader compared to the SW speed jump, and it shows many small-scale changes afterward, as in shock S19. The PUI cooling index is relatively stable throughout the 20-day interval but shows a small decrement around the shock transition. The shock compression ratio based on the PUI density compression is 1.91, and the shock speed in the solar inertial frame is 400 km s$^{-1}$. This is the strongest distant interplanetary shock observed by SWAP so far with the high-resolution data.

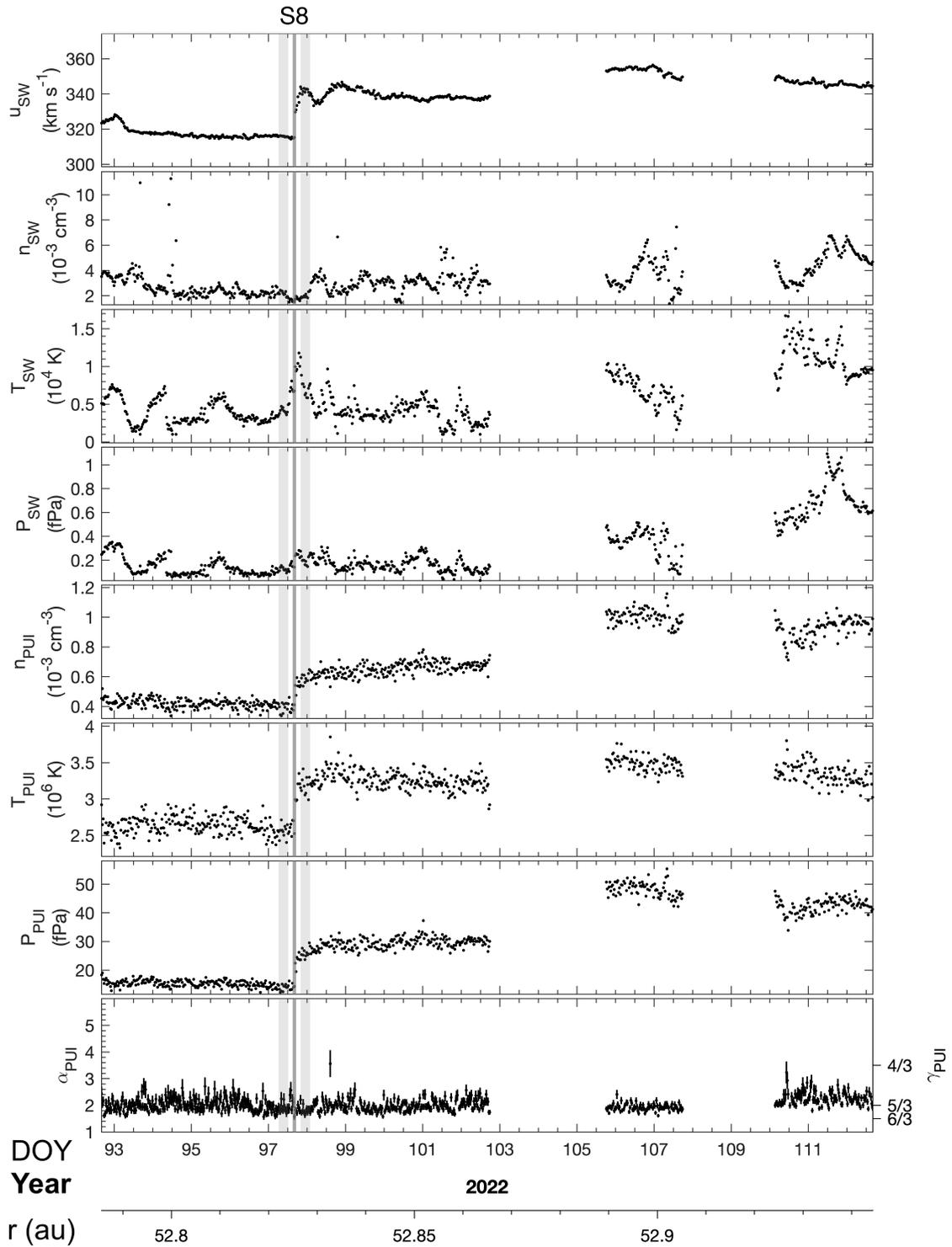

*Figure 2. Variation of SW and PUI properties across shock* S8 *(gray vertical line) observed on 2022 April 07 (DOY 97) at 15:52:35 UTC. The shaded gray regions indicate 6 hr intervals before (upstream) and after (downstream) the shock; average values in these regions are considered as upstream and downstream quantities.*

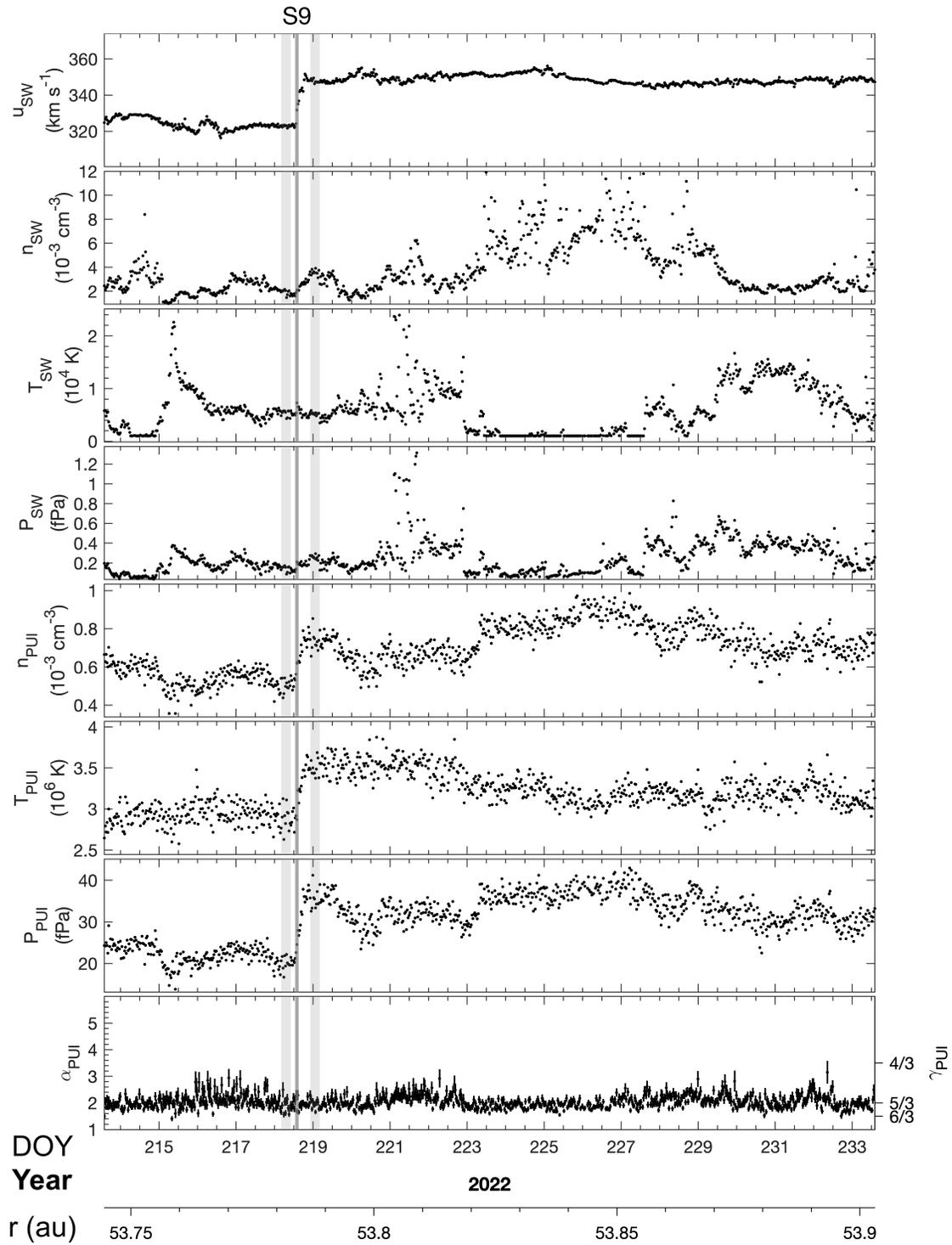

*Figure 3. Same as Figure 2, but for shock S9 observed on 2022 August 06 (DOY 218) at 13:39:15 UTC.*

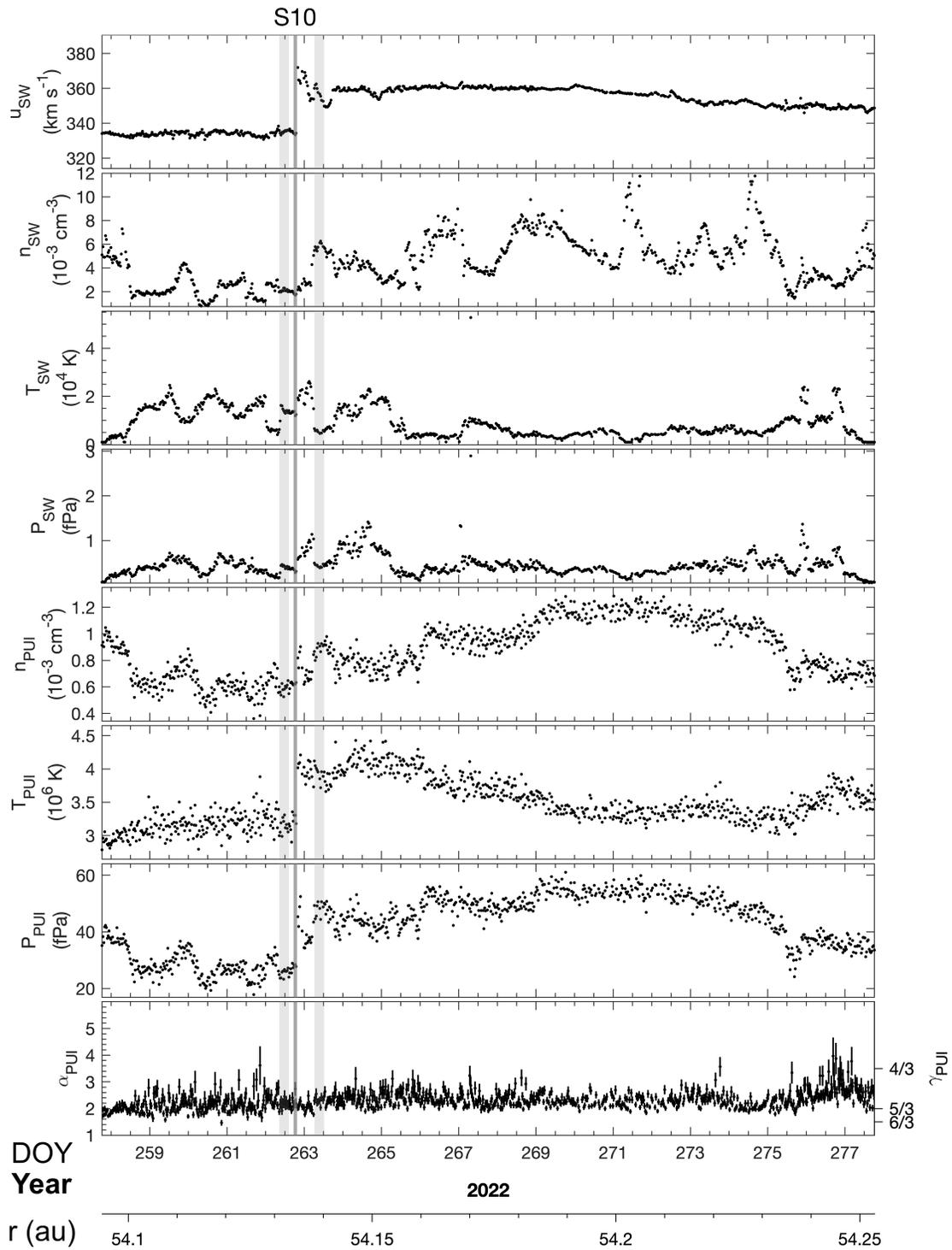

*Figure 4. Same as Figure 2, but for shock S10 observed on 2022 September 19 (DOY 262) at 18:08:03 UTC.*

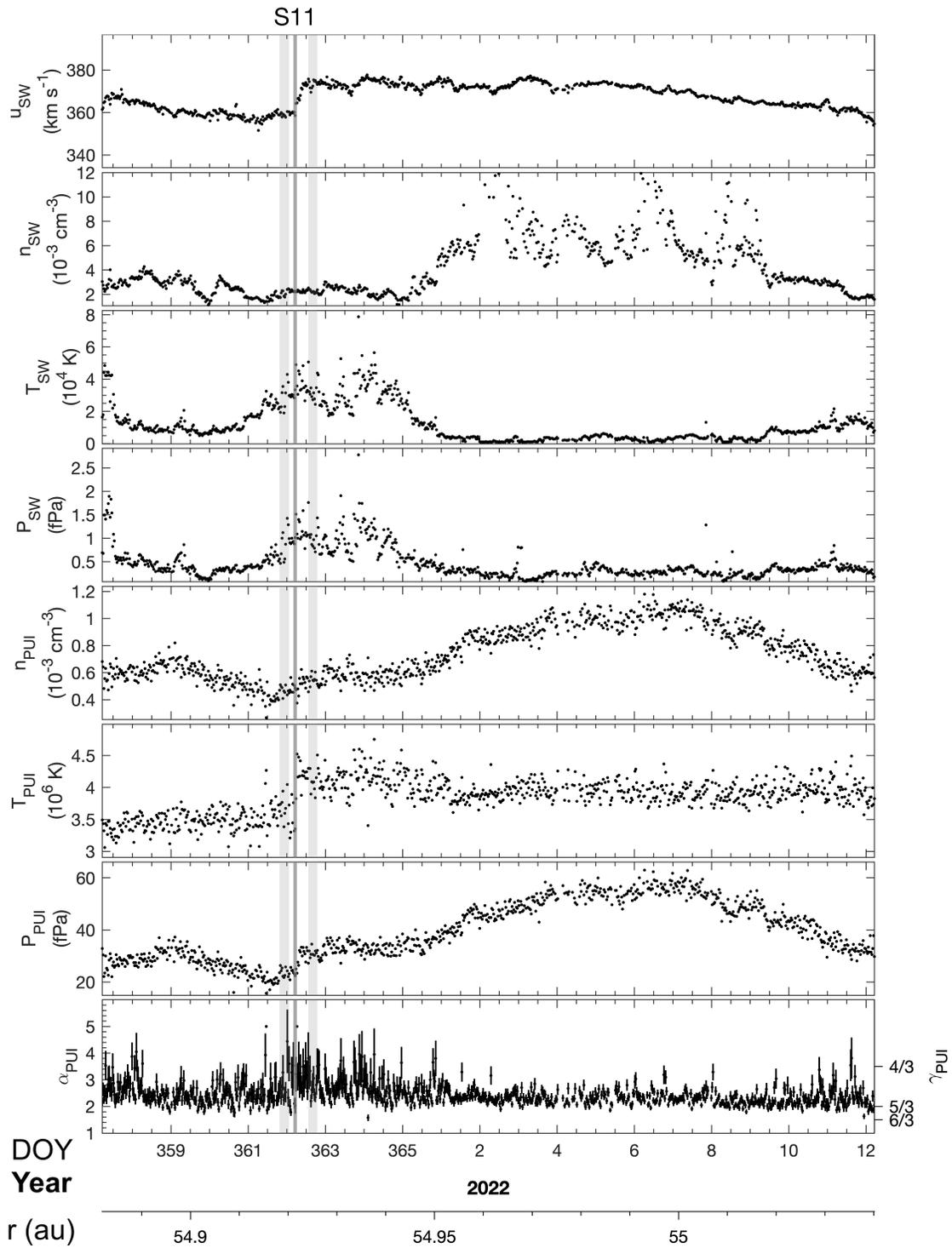

*Figure 5. Same as Figure 2, but for shock S11 observed on 2022 December 28 (DOY 362) at 04:58:43 UTC.*

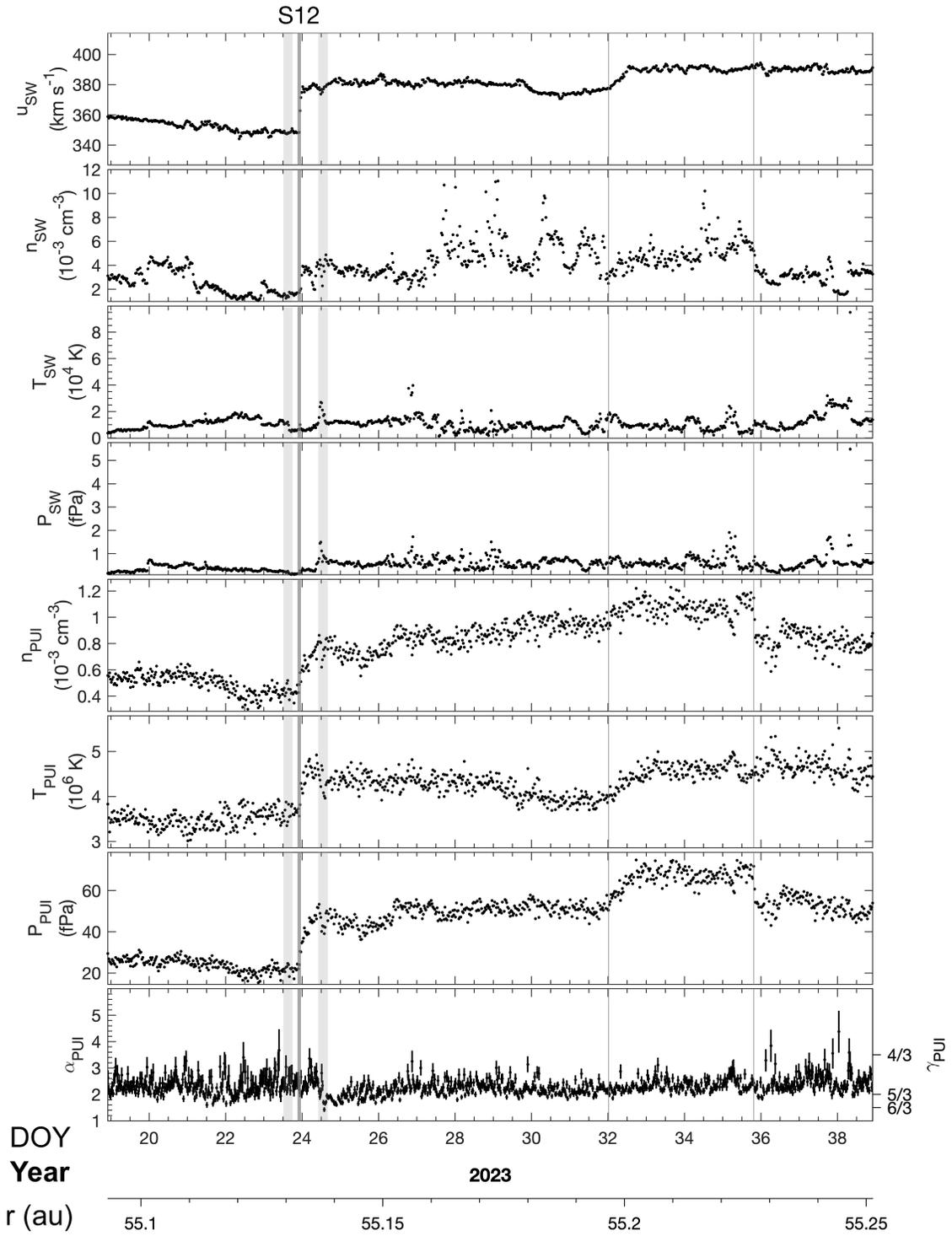

*Figure 6. Same as Figure 2, but for shock* S12 *observed on 2023 January 23 (DOY 23) at 21:57:23 UTC.*

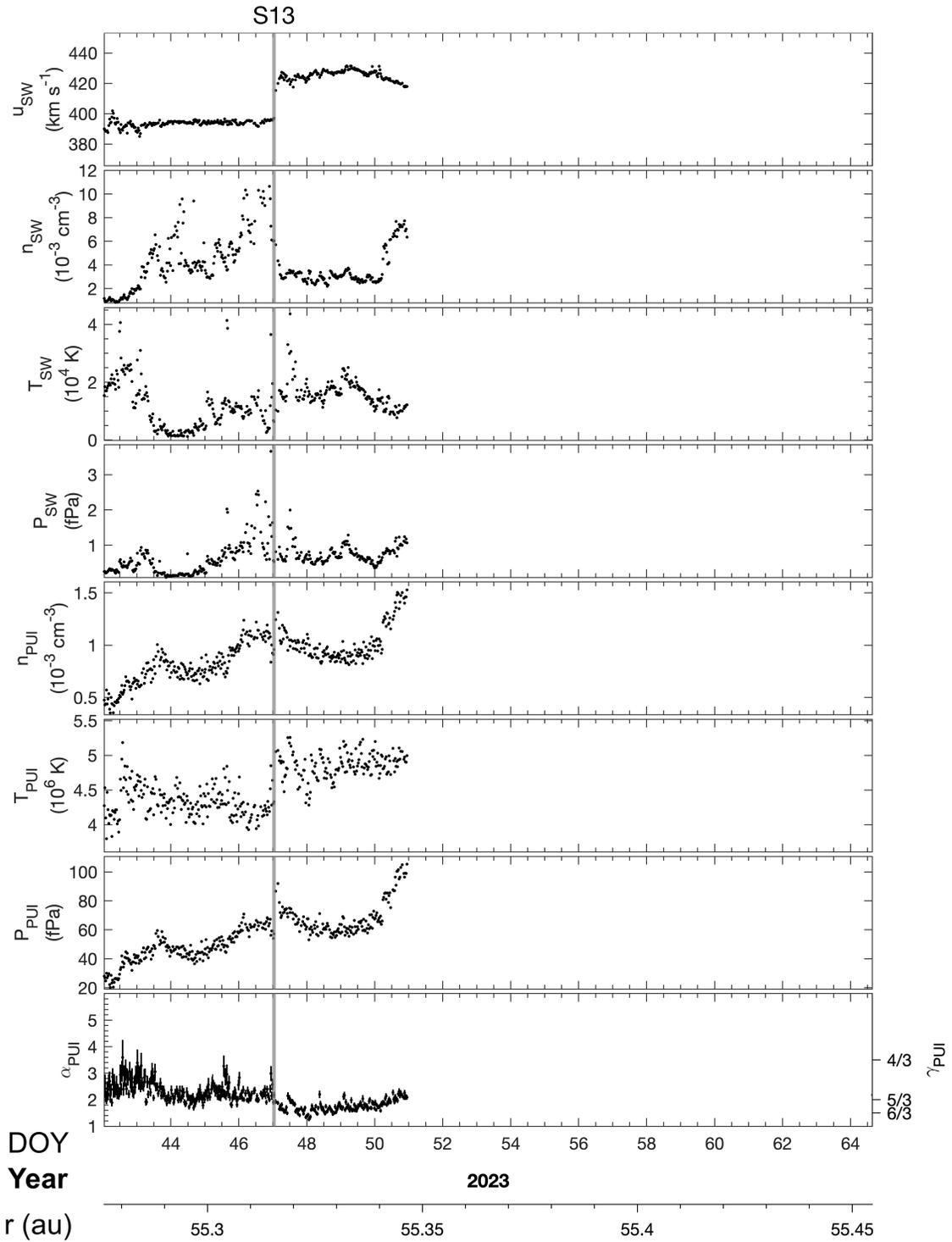

*Figure 7. Same as Figure 2, but for shock S13 observed on 2023 February 16 (DOY 47) at 00:38:27 UTC.*

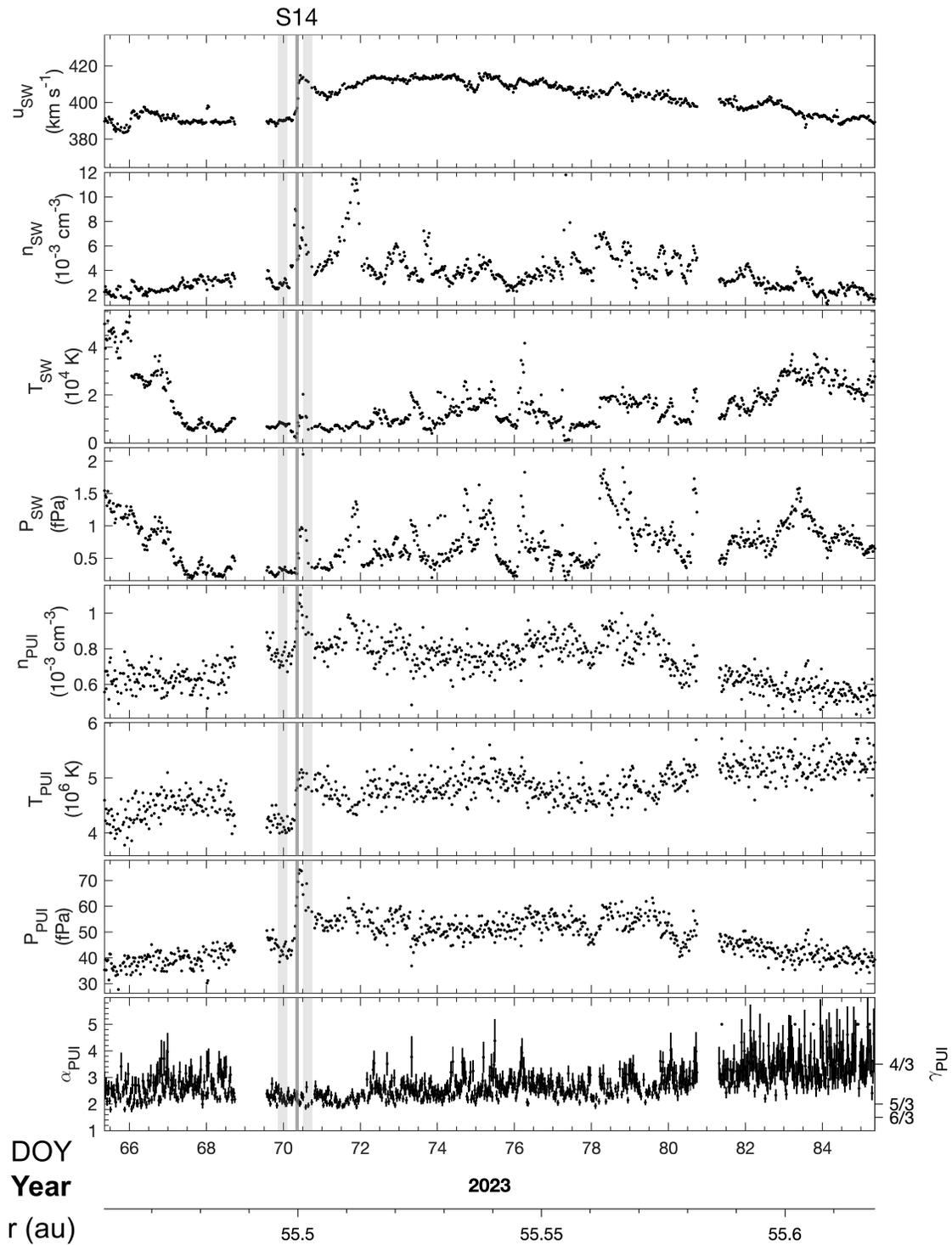

*Figure 8. Same as Figure 2, but for shock S14 observed on 2023 March 11 (DOY 70) at 08:13:55 UTC.*

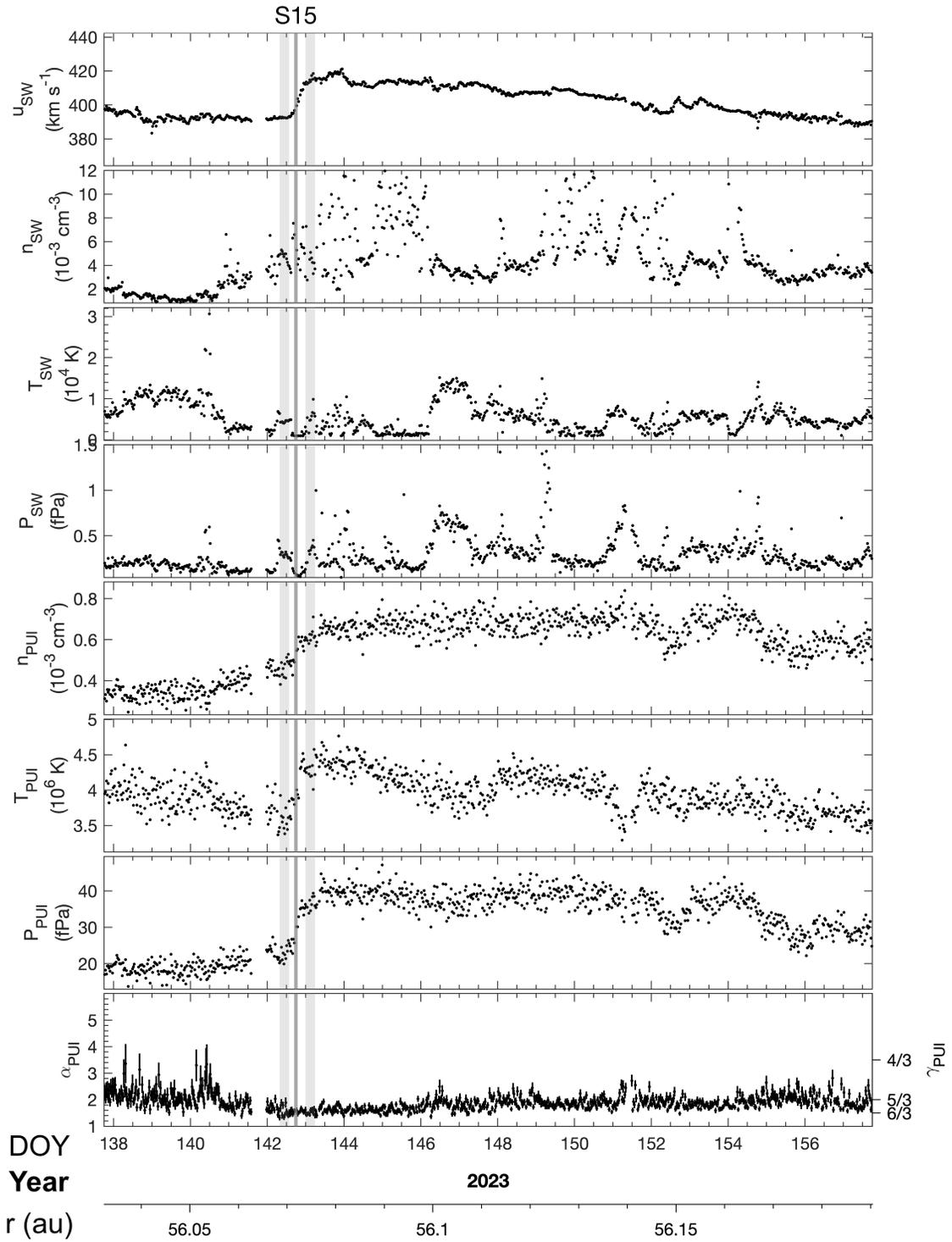

*Figure 9. Same as Figure 2, but for shock S15 observed on 2023 May 22 (DOY 142) at 17:44:35 UTC.*

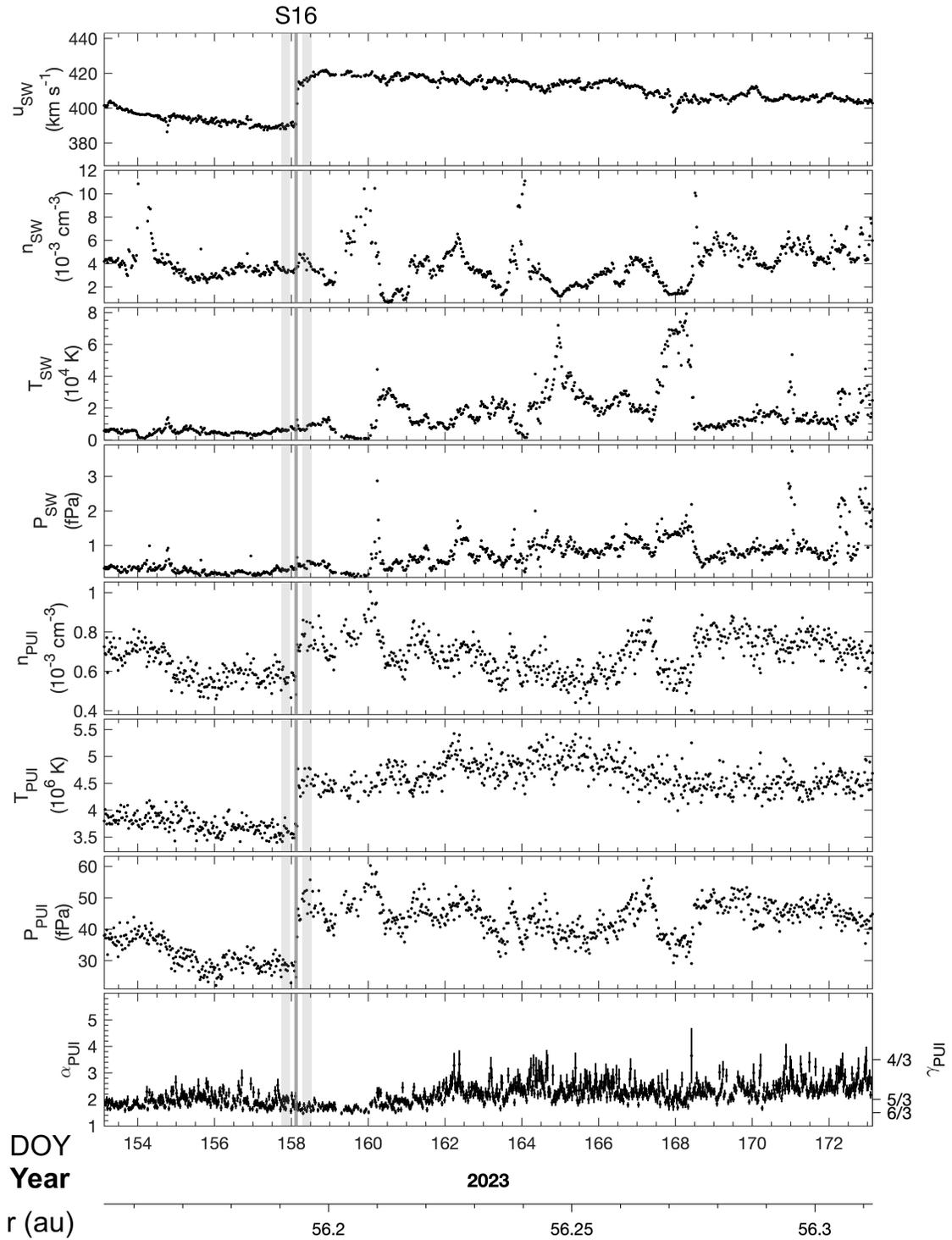

*Figure 10. Same as Figure 2, but for shock S16 observed on 2023 June 07 (DOY 158) at 02:48:35 UTC.*

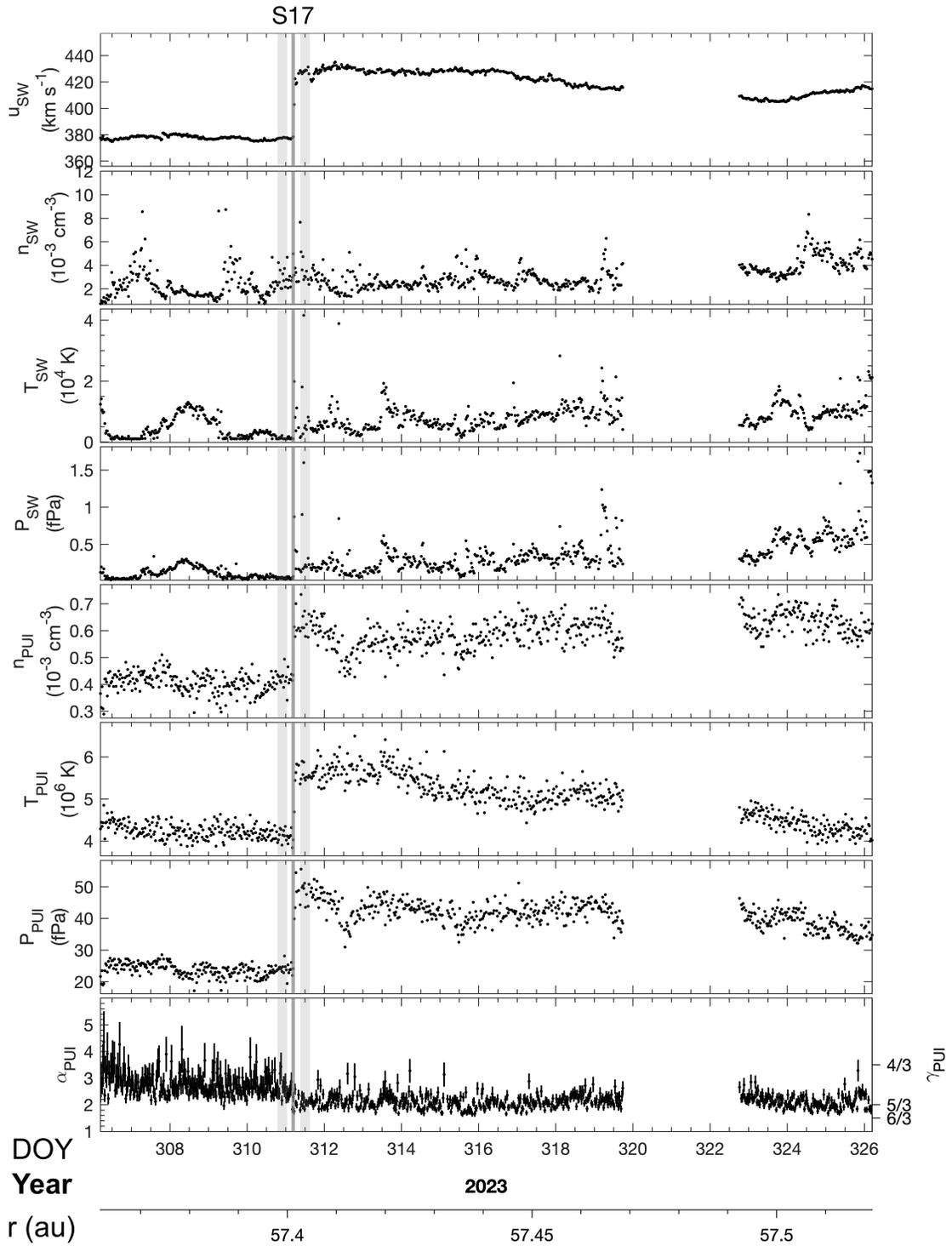

*Figure 11. Same as Figure 2, but for shock S17 observed on 2023 November 07 (DOY 311) at 04:24:35 UTC.*

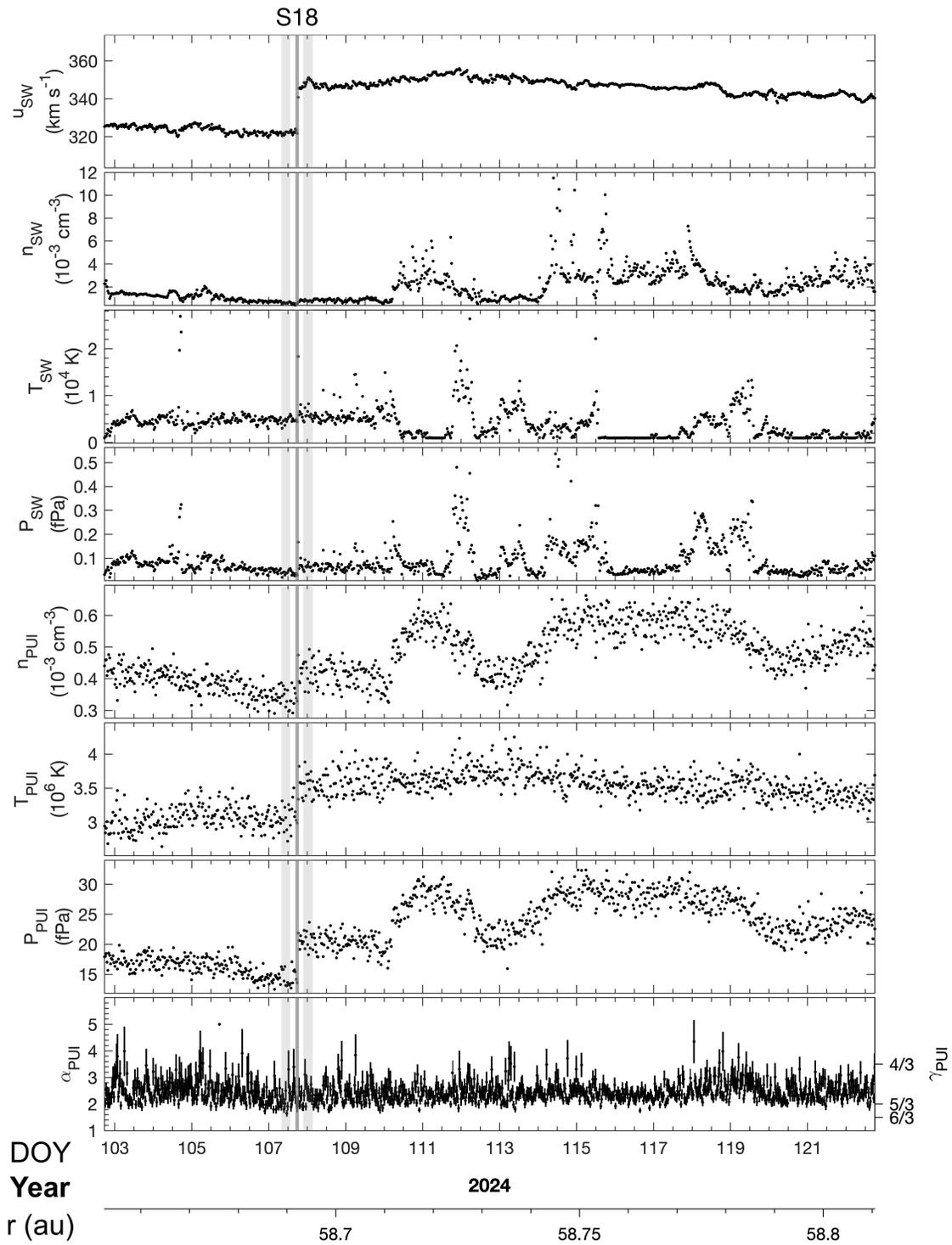

Figure 12. Same as Figure 2, but for shock S18 observed on 2024 April 19 (DOY 107) at 00:38:28 UTC.

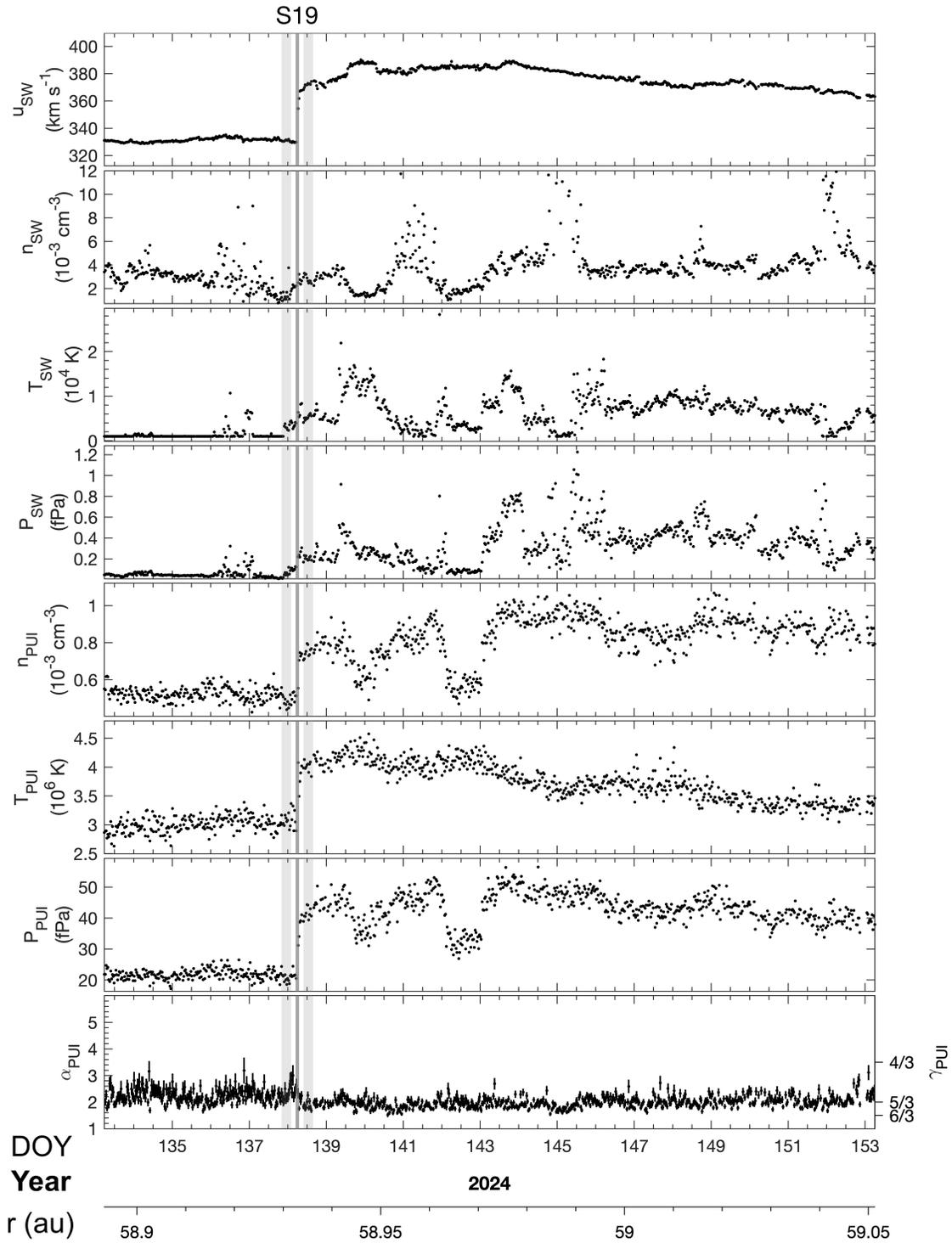

*Figure 13. Same as Figure 2, but for shock* S19 *observed on 2024 May 19 (DOY 138) at 14:47:32 UTC.*

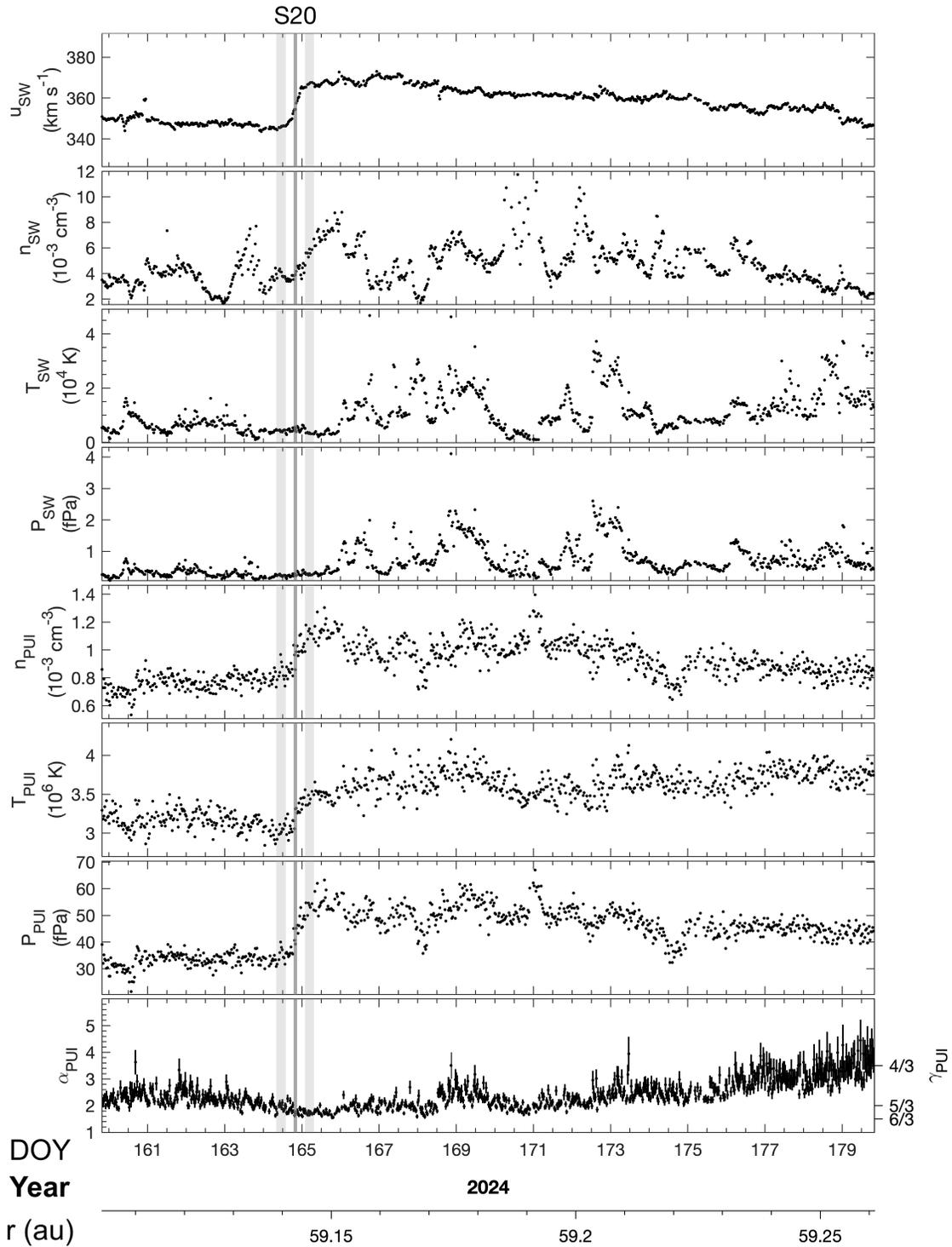

*Figure 14. Same as Figure 2, but for shock S20 observed on 2024 June 15 (DOY 164) at 06:34:44 UTC.*

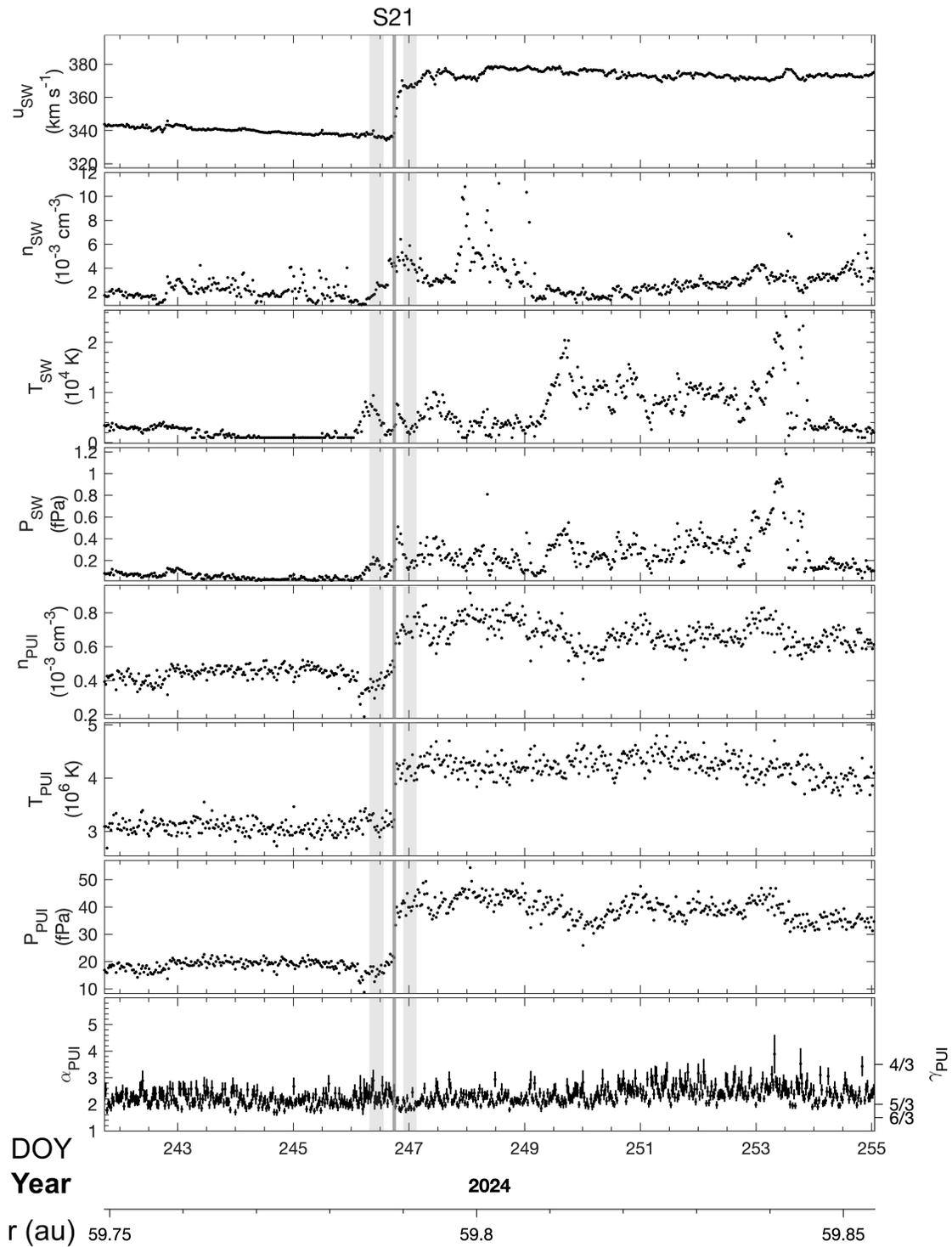

Figure 15. Same as Figure 2, but for shock S21 observed on 2024 September 09 (DOY 246) at 09:49:56 UTC

*Table 1. Upstream (1) and Downstream (2) PUI Parameters, Shock Compression Ratios, and Shock Speeds for all high-resolution shocks observed by New Horizons SWAP. Note that the first seven shocks (1-7) were analyzed by McComas et al. (2022), and the remaining fourteen (8-21) are analyzed in detail in this paper. The values following the ± sign represent the 1-sigma standard error of the mean.*

| Shock | Date | Time (UTC) | Radial Dist. (au) | $n_1$ ($\times 10^{-3}$ cm$^{-3}$) | $n_2$ ($\times 10^{-3}$ cm$^{-3}$) | $u_1$ (km s$^{-1}$) | $u_2$ (km s$^{-1}$) | $r_{\text{comp}}$ | $V_{\text{sh}}$ (km s$^{-1}$) |
|---|---|---|---|---|---|---|---|---|---|
| 1 | 2021-03-22 | 04:58:42 | 49.79 | 0.75 ± 0.01 | 1.04 ± 0.01 | 323.4 ± 0.1 | 333.0 ± 0.1 | 1.39 ± 0.02 | 358 ± 1 |
| 2 | 2021-06-08 | 09:22:11 | 50.41 | 0.68 ± 0.01 | 1.20 ± 0.02 | 318.8 ± 0.1 | 349.7 ± 0.3 | 1.77 ± 0.04 | 390 ± 2 |
| 3 | 2021-08-31 | 04:58:43 | 51.08 | 0.44 ± 0.01 | 0.61 ± 0.01 | 346.9 ± 0.1 | 364.9 ± 0.1 | 1.37 ± 0.04 | 414 ± 5 |
| 4 | 2021-09-09 | 02:48:35 | 51.15 | 0.54 ± 0.01 | 0.67 ± 0.02 | 362.2 ± 0.2 | 372.1 ± 0.2 | 1.24 ± 0.04 | 413 ± 7 |
| 5 | 2021-10-07 | 20:52:19 | 51.38 | -- | -- | -- | -- | -- | -- |
| 6 | 2021-11-09 | 00:38:27 | 51.64 | 0.57 ± 0.01 | 1.00 ± 0.01 | 319.9 ± 0.2 | 357.6 ± 0.5 | 1.77 ± 0.04 | 407 ± 3 |
| 7 | 2021-12-05 | 12:34:11 | 51.85 | 0.68 ± 0.01 | 1.04 ± 0.02 | 348.8 ± 0.1 | 366.7 ± 0.3 | 1.53 ± 0.03 | 401 ± 2 |
| 8 | 2022-04-07 | 15:52:35 | 52.83 | 0.39 ± 0.01 | 0.59 ± 0.01 | 315.9 ± 0.1 | 341.8 ± 0.4 | 1.50 ± 0.04 | 393 ± 5 |
| 9 | 2022-08-06 | 13:39:15 | 53.78 | 0.51 ± 0.01 | 0.74 ± 0.02 | 322.9 ± 0.2 | 347.8 ± 0.3 | 1.45 ± 0.04 | 403 ± 5 |
| 10 | 2022-09-19 | 18:08:03 | 54.13 | 0.59 ± 0.01 | 0.88 ± 0.01 | 335.2 ± 0.3 | 357.0 ± 1.3 | 1.51 ± 0.04 | 400 ± 5 |
| 11 | 2022-12-28 | 04:58:43 | 54.92 | 0.47 ± 0.01 | 0.54 ± 0.01 | 359.2 ± 0.3 | 373.5 ± 0.6 | 1.16 ± 0.04 | 463 ± 21 |
| 12 | 2023-01-23 | 21:57:23 | 55.13 | 0.44 ± 0.01 | 0.77 ± 0.02 | 348.5 ± 0.3 | 377.6 ± 0.5 | 1.74 ± 0.07 | 417 ± 4 |
| 13 | 2023-02-16 | 00:38:27 | 55.32 | -- | -- | -- | -- | -- | -- |
| 14 | 2023-03-11 | 08:13:55 | 55.50 | 0.75 ± | 0.91 ± | 390.2 ± | 411.4 ± | 1.21 ± | 511 ± |

| | | | | 0.01 | 0.02 | 0.2 | 1.0 | 0.03 | 17 |
|---|---|---|---|---|---|---|---|---|---|
| 15 | 2023-05-22 | 17:44:35 | 56.07 | 0.46 ± 0.01 | 0.62 ± 0.01 | 392.6 ± 0.1 | 414.7 ± 0.6 | 1.35 ± 0.04 | 478 ± 8 |
| 16 | 2023-06-07 | 02:48:35 | 56.19 | 0.56 ± 0.01 | 0.79 ± 0.02 | 389.7 ± 0.4 | 416.1 ± 0.6 | 1.40 ± 0.05 | 483 ± 8 |
| 17 | 2023-11-07 | 04:24:35 | 57.40 | 0.42 ± 0.01 | 0.63 ± 0.02 | 377.6 ± 0.1 | 428.2 ± 0.5 | 1.51 ± 0.05 | 528 ± 10 |
| 18 | 2024-04-19 | 00:38:28 | 58.69 | 0.34 ± 0.01 | 0.42 ± 0.01 | 321.9 ± 0.2 | 348.9 ± 0.5 | 1.23 ± 0.05 | 466 ± 25 |
| 19 | 2024-05-19 | 14:47:32 | 58.93 | 0.49 ± 0.01 | 0.75 ± 0.01 | 330.8 ± 0.2 | 371.7 ± 0.5 | 1.54 ± 0.04 | 448 ± 5 |
| 20 | 2024-06-15 | 06:34:44 | 59.14 | 0.83 ± 0.02 | 1.11 ± 0.02 | 345.8 ± 0.2 | 366.7 ± 0.3 | 1.34 ± 0.04 | 429 ± 8 |
| 21 | 2024-09-05 | 09:49:56 | 59.79 | 0.38 ± 0.01 | 0.72 ± 0.01 | 336.8 ± 0.4 | 366.8 ± 0.3 | 1.91 ± 0.07 | 400 ± 3 |

Using these detailed high-resolution observations from SWAP, we are able to study the statistical properties of shock parameters for distant interplanetary shocks over the heliocentric distance range of ~49.5-60 au. For this, we also include the previous 6 fast-forward shocks analyzed by McComas et al. (2022). However, we excluded the fast-reverse shock S5 from McComas et al. (2022) because the SW density appears to be compressed much more than the PUI density (see Figure 8 in McComas et al. (2022)), suggesting a possible different physical mechanism than for the fast-forward shocks. Figure 16 shows the histogram of the shock compression ratio and shock speed in the upstream plasma frame for all 19 fast-forward shocks observed by *New Horizons* SWAP (as mentioned before, shock S13 is excluded from statistical analysis because we could not confidently derive the shock parameters). The upstream and downstream values, shock compression ratios, and shock speeds in the solar inertial frame are listed in Table 1. Note that the shock parameters in Table 1 for shocks 1-7 are slightly different than those reported in McComas et al. (2022) because of narrower intervals (~6 hr) used here for upstream and downstream regions, but the values are within the 1-sigma uncertainty of the values in McComas et al. (2022). The shock compression ratio over this heliocentric distance range as shown in Figure 16 is relatively small and ranges from 1.16 to 1.91, with the mode around 1.3.

The shock speed in the upstream plasma frame ($|V_{sh} - u_1|$) ranges from 34.2 – 151 km s$^{-1}$, with the mode around 60 km s$^{-1}$.

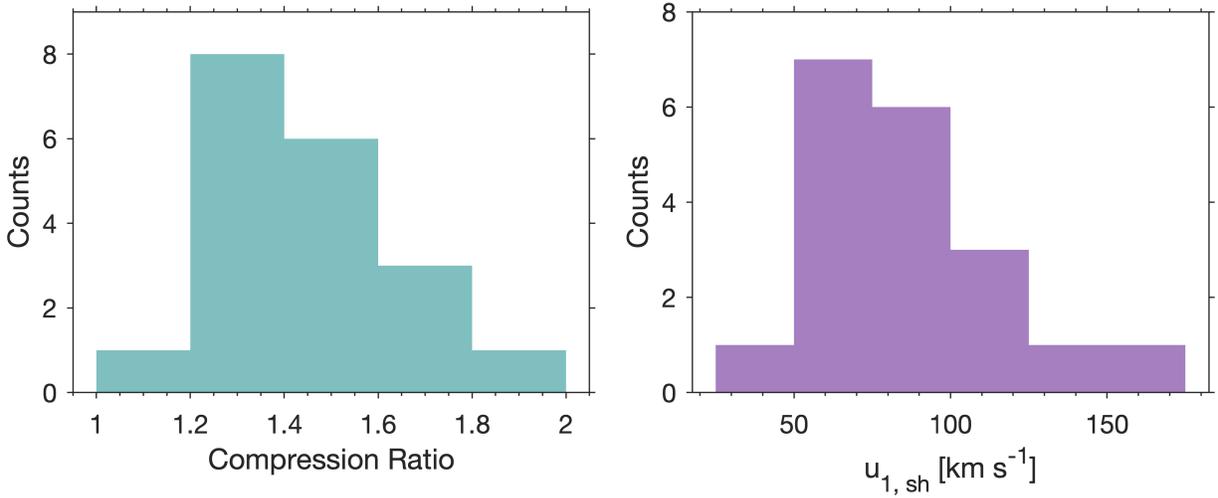

*Figure 16. Histogram of (left panel) shock compression ratio, (right panel) shock speed in the upstream plasma frame.*

Figure 17 shows the variation of SW density compression $\left(\frac{n_{2,SW}}{n_{1,SW}}\right)$ from upstream to downstream with the PUI density compression. We can see that the SW density compression is weakly correlated with the PUI density compression with a Pearson correlation coefficient of only 0.31. This means that a stronger shock is slightly more likely to show a compression in the SW density downstream, though the correlation coefficient is not strong enough to derive a definitive conclusion. Perhaps more, future data from SWAP will enable us to test this hypothesis.

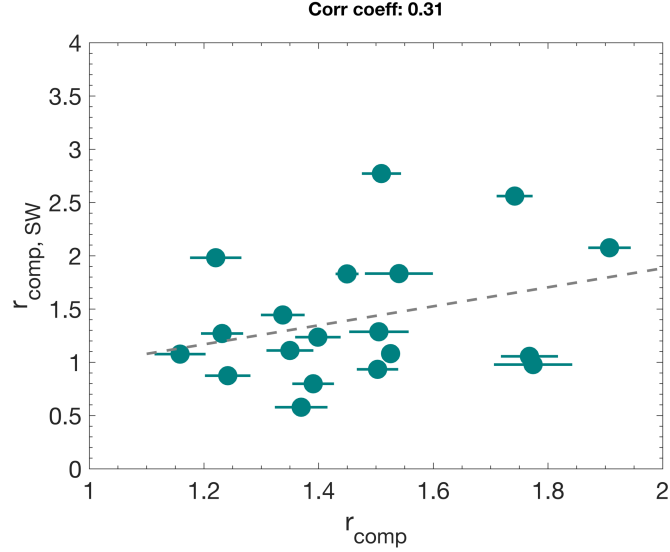

*Figure 17. Variation of SW density compression from upstream to downstream with the PUI density compression. The horizontal and vertical error bars represent the propagated 1-sigma standard error of the mean. The dashed gray line represents the linear least-square line, and the value on the top represents the Pearson correlation coefficient.*

### 3.2 Sonic Mach Number

Usually, in the SW from ~1-10 au, the fast magnetosonic speed is the critical speed for defining shocks as both the SW plasma and magnetic field pressures are significant and comparable to each other. However, in the outer heliosphere, the PUI thermal pressure rapidly becomes dominant, and the critical speed for shocks collapses back to the PUI sound speed. Generally, the sonic Mach number ($M_c$) is defined as the ratio of the upstream/downstream plasma flow speed in the shock frame to the local sound speed in the plasma, given by

$$M_{c,i} = \frac{u_{i,\text{sh}}}{c_i}, \quad (3)$$

where $u_{i,\text{sh}}$ is the upstream/downstream plasma flow speed in the shock frame, $c_i$ is the sound speed given by $c_i = \sqrt{\gamma P_i / \rho_i}$. Here, $P_i$ is the total pressure, which is dominated by PUIs, is $P = P_{SW} + P_{PUI} + P_{He^{2+}}$, and $\rho_i$ is the total mass density ($\rho = m_p n_{SW} + m_p n_{PUI} + m_{He^{2+}} n_{He^{2+}}$), $i = 1, 2$ representing upstream and downstream regions, respectively, and $\gamma$ is the polytropic index. For simplicity, we consider adiabatic polytropic index, i.e., $\gamma = \frac{5}{3}$ for all species. Typically, the SW protons, PUIs, and SW alphas have different polytropic indices (Elliott et al. 2019;

Livadiotis et al. 2024) which are rather difficult to obtain for each 30-minute resolution SWAP data.

The variation of the sonic Mach number downstream with the upstream value is shown in Figure 18. The sonic Mach numbers are color-coded by the difference in the plasma flow speed in the shock frame between upstream and downstream regions. Out of 19 high-resolution shocks/waves observed by SWAP, 11 have an upstream sonic Mach number greater than one, i.e., the upstream plasma flow is supersonic. The remaining 8 have upstream sonic Mach numbers less than or equal to one, indicating their subsonic nature, suggesting they are already degraded into compressional waves. Moreover, in the newer data, ~64% of the shocks are supersonic (9 out of 14). For the shocks with an upstream sonic Mach number greater than 1, the minimum value of the difference in the flow speed between upstream and downstream is 14.2 km s$^{-1}$. In addition, out of 11 shocks with supersonic upstream flow, 9 have transitioned to subsonic in the downstream region. For shocks S14 and S17 (newer shocks), the downstream sonic Mach number is less than 1 only while considering their 1-sigma uncertainty. Finally, the sonic Mach numbers for all shocks/waves are reduced across the shock. The upstream and downstream sonic Mach numbers for all 19 shocks are also listed in Table 2.

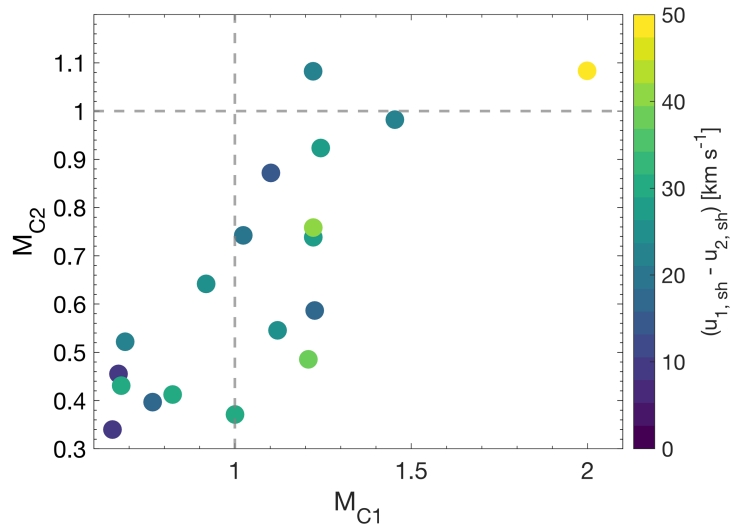

*Figure 18. Variation of the downstream sonic Mach number with the upstream sonic Mach number. The sonic Mach numbers are color-coded by the difference in the plasma flow speed in the shock frame between upstream and downstream regions.*

3.3    Theoretical Shock Compression Ratio

The theoretical shock compression ratio from the Rankine-Hugoniot (RH) relations, assuming that shocks are exactly perpendicular, can be obtained by solving the shock adiabatic of the following form:

$$2(2-\gamma)R^2 + (2\beta_1 + (\gamma-1)\beta_1 M_{C1}^2 + 2)\gamma R - \gamma(\gamma+1)\beta_1 M_{C1}^2 = 0,$$

where $\gamma$ is the polytropic index (assumed constant for both upstream and downstream regions for all species), $\beta_1$ is the plasma beta upstream of the shock ($\beta_1 = P_1 / \frac{B_1^2}{2\mu_0}$, $B_1$ is the upstream magnetic field and $\mu_0$ is the permeability of free space), $M_{C1}$ is the upstream sonic Mach number, and $R$ is the theoretical shock compression ratio. Because the magnetic field measurements during New Horizons' shock crossing were missing, we estimate the magnetic field upstream of the shock by propagating the magnetic field from 1 au to the shock location using Parker spiral equations. The timing of the magnetic field data at 1 au is chosen by backtracking the SW flow from the observed shock location to 1 au using the SW speed observed by SWAP. The slowing of SW by mass-loading is accounted for using a first-order approximation of ~7% slowdown from Elliott et al. (2019). A comparison of the theoretical shock compression ratios derived from the RH relations and observed PUI density compression is shown below in Figure 19.

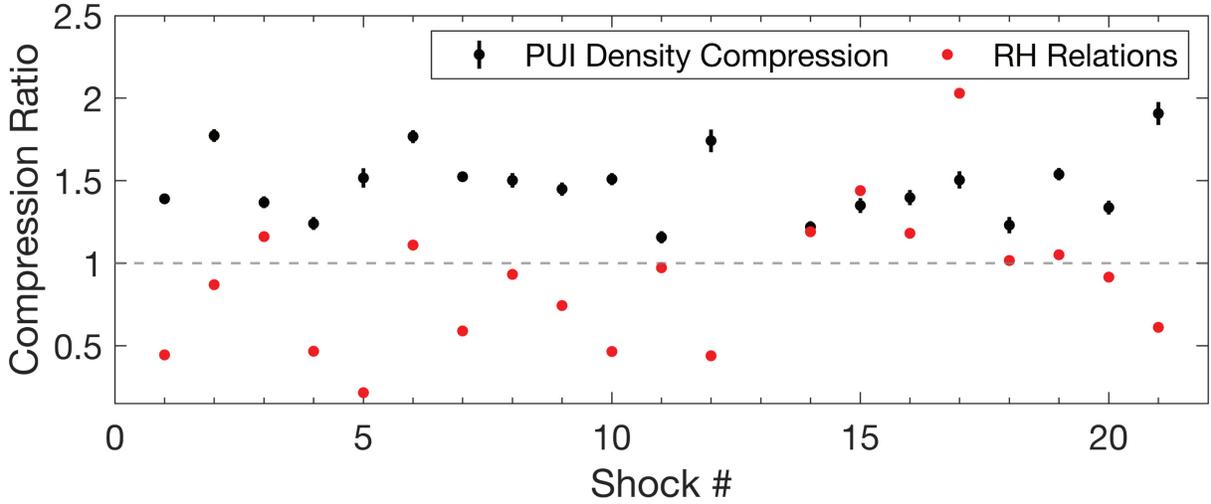

*Figure 19. Comparison of the shock compression ratio obtained from PUI density compression (black dots with error bars) and theoretical Rankine-Hugoniot relations (red dots). The error bars represent 1-sigma standard error of the mean.*

As one can see, the theoretical shock compression from RH relations is very close to the PUI density compression for shocks S14 and S15, close for S3, S11, and S16, but quite different for other shocks. The theoretical compression ratios are often smaller than one, which occurs when the upstream Sonic Mach numbers are close to or less than one (see Table 2). Note that the shock adiabatic in the above Equation is very simplified, and a more accurate theoretical calculation requires the in-situ magnetic field measurements, polytropic behavior, and possible coupling with different species (e.g., Livadiotis 2015), which are generally unknown across these shocks.

3.4     Shock Transition Scale

The classical structure of the quasi-perpendicular shock contains a foot, ramp, and overshoot (Treumann 2009). The foot is formed by the reflection of PUIs by the electrostatic cross-shock potential or magnetic field pileup ahead of the shock, causing the upstream bulk SW ions to slow down ahead of the shock (Kumar et al. 2018). Because *New Horizons* does not carry a magnetometer and thus does not have magnetic field measurements, it is difficult to identify the detailed shock structure and accurately estimate the shock width. Therefore, we use the PUI density profile between upstream and downstream regions to calculate the shock transition scale (shock width). For this, we first transform the SWAP observations in the shock rest frame, where the SW flow speed in the shock frame is given as $u_{SW,sh} = |V_{sh} - u_{SW}|$. *Figure 20* shows an example of the SW speed (top panel) and PUI density (bottom panel) in the shock frame for shock S9. The time and distance are measured from the shock observation time and the shock position, respectively. In the shock frame, a decrease in the SW speed is observed from the upstream to the downstream region. The energy equivalent to the speed difference is utilized to heat particles in the downstream region, most of which goes to the PUIs (see Section 3.4 for details).

We estimate the shock transition scale by fitting a tangent hyperbolic function to the PUI density profile between upstream and downstream regions (see bottom panel of *Figure 20*). The fitted tangent hyperbolic function has the form

$$y = a \tanh(k(x - x_0)) + b, \qquad (4)$$

where $a$ is the amplitude that determines the PUI density jump, $b$ represents the midpoint of the hyperbolic curve between upstream and downstream value, $k$ is the slope that characterizes the shock transition scale, and $x_0$ represents the shock position. We define the shock transition scale as a distance between two points in the shock frame, where the PUI density rises from 16% to 84% (orange-colored vertical dashed lines) of the full density jump in the fitted hyperbolic function (solid green curve). The shock transition scale for shock S9 in the shock frame is ~0.036 au.

We also represent the shock transition scale in terms of an average, estimated upstream PUI advective gyroradius, $r_g$. For this, we estimate the magnetic field upstream of the shock using the magnetic field propagated from 1 au to the shock location using Parker spiral equations (see Section 3.3). This gives the shock transition scale for shock S9 as ~160 $r_g$ or ~$5.4 \times 10^6$ km. The shock transition scale for all 19 fast-forward shocks observed by *New Horizons* SWAP is listed in Table 2. Note that these shocks are much wider than shocks observed at 1 au and the HTS.

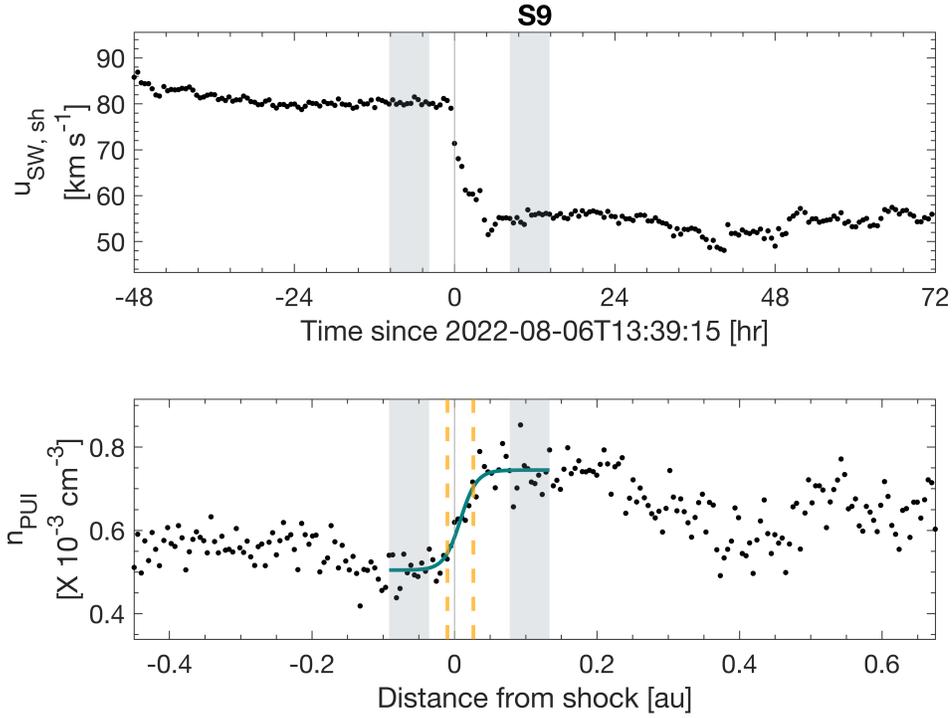

*Figure 20. (Top panel) SW bulk flow speed in the shock frame and (bottom panel) PUI density profile across the shock S9. The vertical solid line represents the position of the shock, and the two vertical orange-dashed lines represent the distances in the shock frame where PUI density increases by 16% and 84% of the full density jump from upstream to downstream.*

Now, we study how the shock transition scale varies with different shock parameters. Figure 21 shows the variation of the shock transition scale with the compression ratio, shock speed in the upstream plasma frame, and the upstream sonic Mach number. The shock width appears independent of the shock compression ratio (left panel), indicated by a very small Pearson correlation coefficient of 0.05. A similar result was reported by Zirnstein et al. (2023) from the PIC simulations of quasi-perpendicular HTS. On the other hand, the shock width decreases with increasing shock speed in the upstream plasma frame (middle panel), indicated by a correlation coefficient of -0.60. In addition, the shock width also decreases with increasing upstream sonic Mach number (right panel), though the correlation is weak (-0.27).

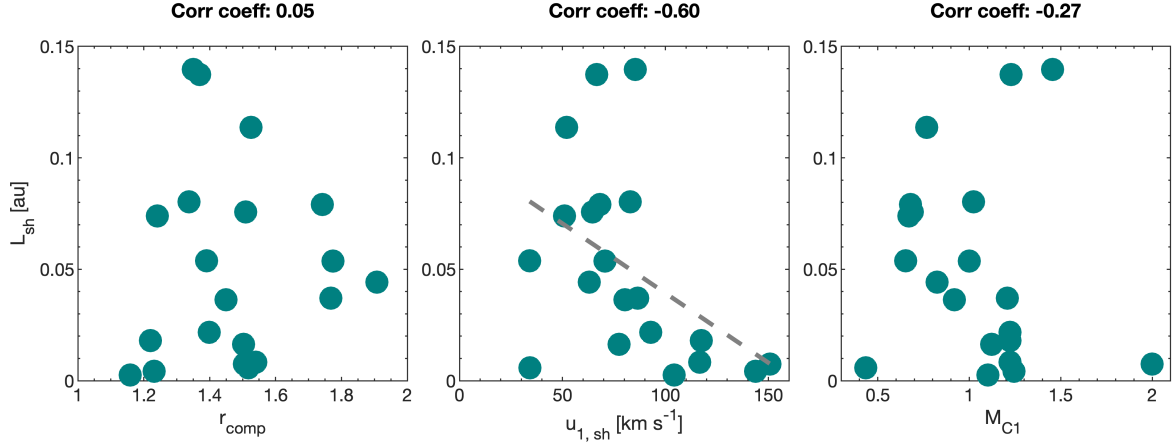

*Figure 21. Variation of the shock transition scale with (left panel) the compression ratio, (middle panel) shock speed in the upstream plasma frame, and (right panel) upstream sonic Mach number. The gray dashed lines represent the linear least-square line, and the values on the top represent the Pearson correlation coefficient.*

*Table 2. The upstream and downstream plasma flow speed in the shock frame, sonic Mach number upstream and downstream of the shock in the shock rest frame, and the shock transition scale in two different units (au and upstream PUI advective gyroradius). The values following the ± sign represent the 1-sigma standard error of the mean.*

| Shock | $u_{1,sh}$ [km s$^{-1}$] | $u_{2,sh}$ [km s$^{-1}$] | Sonic Mach # (upstream) $M_{c1}$ | Sonic Mach # (downstream) $M_{c2}$ | Shock Transition Scale [au] | [$r_g$] |
|---|---|---|---|---|---|---|
| 1 | 34.0 ± 1.4 | 24.5 ± 1.4 | 0.65 ± 0.03 | 0.34 ± 0.02 | 0.054 | 220 |
| 2 | 70.7 ± 2.1 | 39.8 ± 2.1 | 1.00 ± 0.04 | 0.37 ± 0.02 | 0.054 | 250 |
| 3 | 66.6 ± 5.0 | 48.6 ± 5.0 | 1.23 ± 0.11 | 0.59 ± 0.06 | 0.137 | 530 |
| 4 | 50.9 ± 7.0 | 41.0 ± 7.0 | 0.67 ± 0.09 | 0.46 ± 0.08 | 0.074 | 260 |
| 5 | -- | -- | -- | -- | -- | -- |
| 6 | 86.6 ± 2.8 | 49.0 ± 2.8 | 1.21 ± 0.05 | 0.49 ± 0.06 | 0.037 | 190 |
| 7 | 52.0 ± 2.2 | 34.1 ± 2.3 | 0.77 ± 0.07 | 0.40 ± 0.03 | 0.114 | 430 |
| 8 | 77.4 ± 5.0 | 51.5 ± 5.0 | 1.12 ± 0.07 | 0.55 ± 0.03 | 0.016 | 79 |
| 9 | 80.3 ± 5.0 | 55.4 ± 5.0 | 0.92 ± 0.06 | 0.64 ± 0.05 | 0.036 | 160 |
| 10 | 64.5 ± 5.0 | 42.8 ± 5.0 | 0.69 ± 0.05 | 0.52 ± 0.06 | 0.076 | 360 |
| 11 | 104.2 ± 21.0 | 89.9 ± 21.0 | 1.10 ± 0.23 | 0.87 ± 0.21 | 0.003 | 12 |
| 12 | 68.2 ± 4.0 | 39.1 ± 4.0 | 0.68 ± 0.04 | 0.43 ± 0.04 | 0.079 | 360 |
| 13 | -- | -- | -- | -- | -- | -- |

| 14 | 117.4 ± 16.0 | 96.2 ± 16.0 | 1.22 ± 0.17 | 1.10 ± 0.18 | 0.018 | 80 |
|---|---|---|---|---|---|---|
| 15 | 85.4 ± 9.0 | 63.3 ± 9.0 | 1.45 ± 0.15 | 0.98 ± 0.17 | 0.140 | 550 |
| 16 | 92.8 ± 8.0 | 66.4 ± 8.0 | 1.22 ± 0.11 | 0.74 ± 0.09 | 0.022 | 84 |
| 17 | 150.8 ± 10.0 | 100.2 ± 10.0 | 2.00 ± 0.15 | 1.10 ± 0.14 | 0.008 | 32 |
| 18 | 143.8 ± 25.0 | 116.7 ± 25.0 | 1.24 ± 0.22 | 0.92 ± 0.20 | 0.004 | 24 |
| 19 | 116.7 ± 5.0 | 75.8 ± 5.0 | 1.20 ± 0.08 | 0.76 ± 0.06 | 0.008 | 47 |
| 20 | 83.0 ± 8.0 | 62.0 ± 8.0 | 1.02 ± 0.10 | 0.74 ± 0.10 | 0.080 | 380 |
| 21 | 63.0 ± 2.6 | 33.0 ± 2.6 | 0.82 ± 0.04 | 0.41 ± 0.03 | 0.044 | 180 |

## 3.5  Energy Flux Across Distant Interplanetary Shock

To quantize the amount of energy gained by PUIs across the distant interplanetary shock, we calculate the energy density flux for each particle species using MHD energy conservation across a perpendicular shock. The energy density flux ($E_i$) for each species '$i$' is given by (note that we have assumed the shocks are perpendicular)

$$E_i = \left(\frac{1}{2} m_i n_i u_{\text{sh}}^2 + \frac{\gamma}{\gamma - 1} n_i k_B T_i\right) u_{\text{sh}}, \quad (5)$$

where $m_i$ is the ion mass, $n_i$ is the number density, $u_{\text{sh}}$ is the SW flow speed in the shock frame ($u_{\text{sh}} = |V_{\text{sh}} - u_{\text{SW}}|$), $\gamma$ is the polytropic index, $k_B$ is the Boltzmann constant, and $T_i$ is the ion temperature. We use $\gamma = \frac{5}{3}$ as an overall polytropic index for all species as done by Zirnstein et al. (2018) and McComas et al. (2022). Note that this value may vary because of its connection with the cooling index (Livadiotis et al. 2024). The first and second terms in Equation (5) represent the contribution of the dynamic and thermal energy to the energy flux. Equation (5) does not include the contribution of magnetic energy flux. The magnetic field was not measured across these shocks because of the lack of a magnetometer on *New Horizons*. Moreover, Zirnstein et al. (2018) reported that the magnetic field holds a small amount of energy flux downstream of a strong interplanetary shock at a heliocentric distance of ~34 au. The magnetic energy flux will be even smaller over the distance range considered in this study because $B \propto r^{-1}$. Therefore, we do not include it in the total energy flux.

The variation of energy flux for SW and PUIs across all 13 shocks analyzed in detail in this study is shown in Figure 22. In all shocks, the energy flux of PUIs is much higher than SW in the upstream region. In general, the SW energy flux decreases after the shock passage (except for shock S14), while the PUI energy flux increases. This is expected because the shock extracts the energy from the SW flow and converts it to the PUI thermal energy. Shock S14 is quite different, where the SW energy flux downstream is slightly larger than the upstream value. In addition, the SW consists of a significant fraction of energy flux, both upstream and downstream, compared to other shocks. The unusual behavior of SW mass flux across the shock S14 is possibly caused by challenges in accurately identifying the upstream and downstream regions due to a strange speed profile across the shock transition (see Figure 8). In addition, considering the 1-sigma uncertainty of the difference in SW energy flux (see Table 3 and Figure 23), SW can still be losing energy flux downstream. We have also calculated the alpha particle ($He^{2+}$) energy flux using SWAP count rates around the alpha peak, assuming they comove with the SW. The energy flux for SW, PUIs, and alpha particles for both upstream and downstream for all 19 shocks is listed in Table 3.

Figure 23 shows the variation of the difference in the SW and PUI energy flux with the (i) shock compression ratio (panel (a)) and (ii) shock speed in the upstream plasma frame (panel (b)). The difference in the energy flux for SW and PUIs are calculated as, $\Delta E_i = E_{i,2} - E_{i,1}$, where $E_{i,1}$ and $E_{i,2}$ are energy flux upstream and downstream, respectively, taken as average values over the shaded regions of ~6 hr interval in Figure 22. In general, as the shock compression ratio increases, SW ions lose more energy flux downstream; consequently, PUIs gain more energy flux. This behavior in the energy flux of SW and PUIs is more pronounced with the shock speed in the upstream plasma frame.

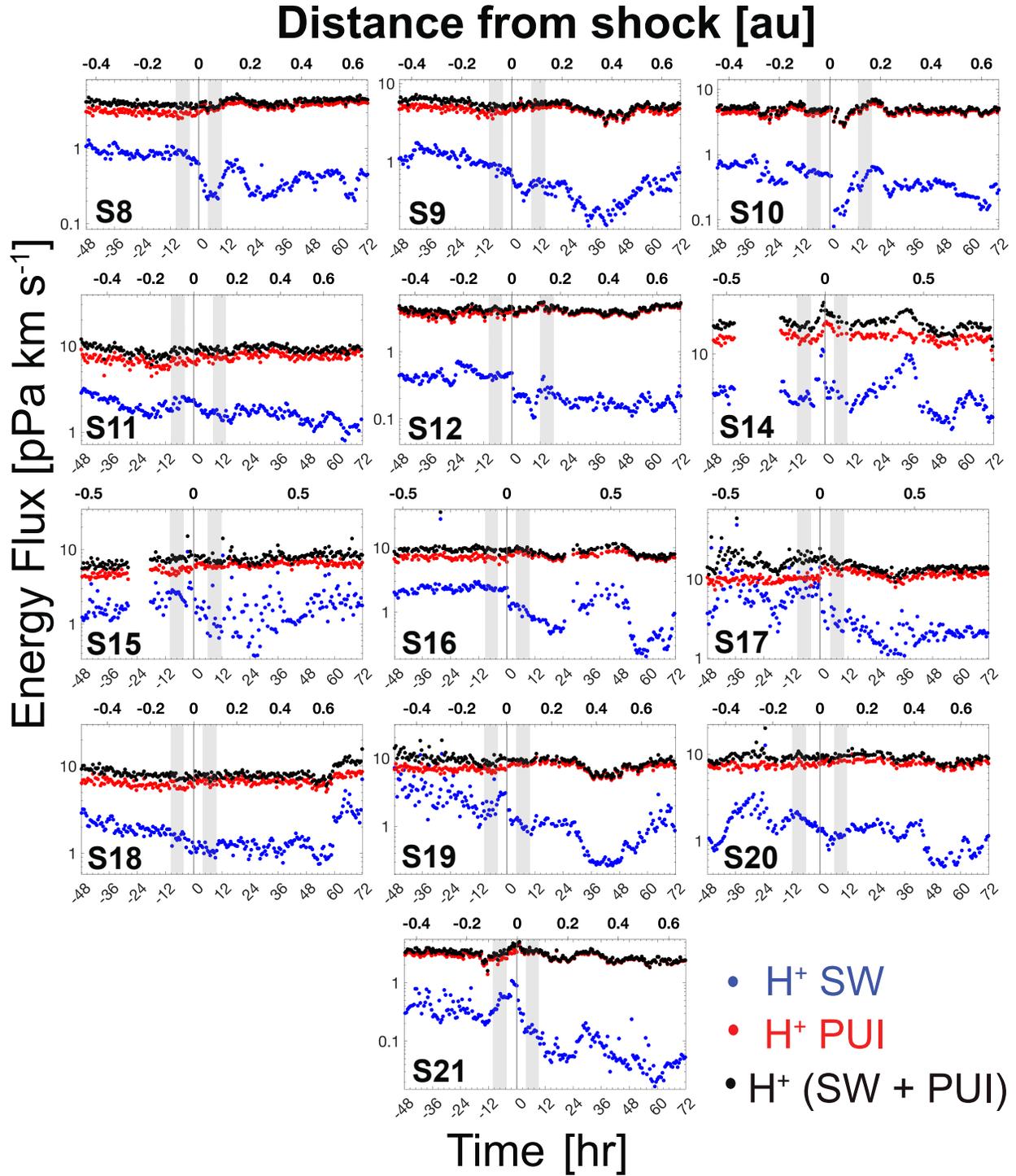

*Figure 22. The energy flux of SW and PUIs across 13 new high-resolution shocks. The shaded gray areas indicate the upstream and downstream regions used to calculate the values shown in Table 3.*

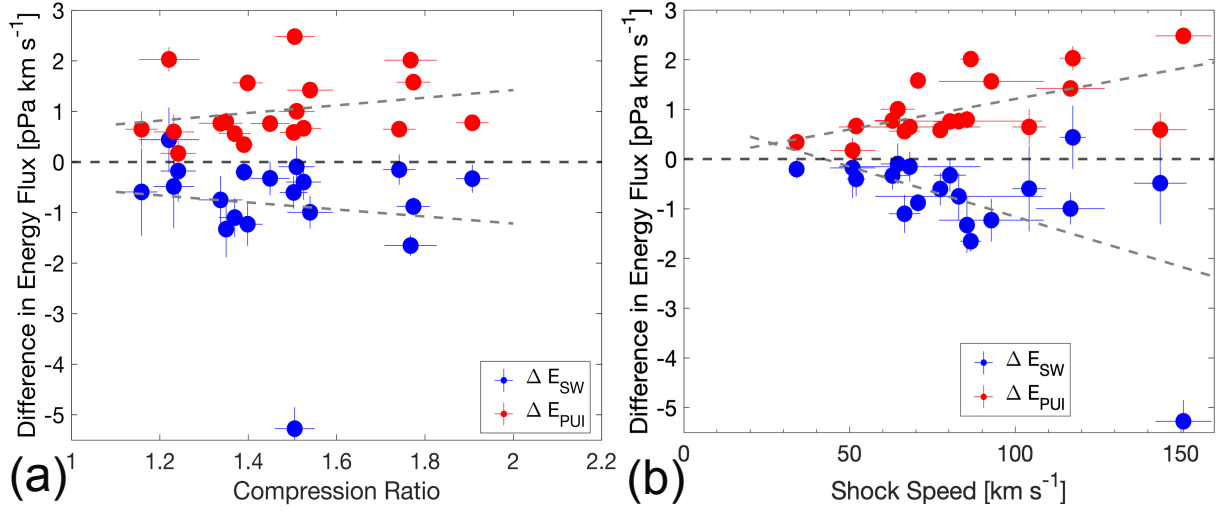

*Figure 23. Variation of difference in the SW and PUI energy flux across the shock with (panel (a)) shock compression ratio and (panel (b)) shock speed in the upstream plasma frame. The vertical and horizontal error bars represent the 1-sigma standard error of the mean. The dashed gray lines represent linear least-square lines.*

*Table 3. Energy Flux upstream (1) and downstream (2) are in units of pPa km s$^{-1}$ for all species. The values following the $\pm$ sign represent the 1-sigma standard error of the mean.*

| **Shock** | $E_{SW,1}$ | $E_{SW,2}$ | $E_{PUI,1}$ | $E_{PUI,2}$ | $E_{alpha,1}$ | $E_{alpha,2}$ | $E_{tot,1}$ | $E_{tot,2}$ |
|---|---|---|---|---|---|---|---|---|
| **1** | 0.32±0.04 | 0.11±0.02 | 2.35±0.10 | 2.69±0.16 | 0.01±0.01 | 0.01±0.01 | 2.67±0.11 | 2.80±0.16 |
| **2** | 1.13±0.11 | 0.25±0.04 | 4.59±0.18 | 6.17±0.37 | 0.14±0.06 | 0.03±0.01 | 5.87±0.22 | 6.46±0.37 |
| **3** | 1.44±0.37 | 0.34±0.10 | 3.26±0.27 | 3.82±0.39 | 0.07±0.01 | 0.06±0.02 | 4.76±0.46 | 4.22±0.41 |
| **4** | 0.36±0.14 | 0.18±0.08 | 3.27±0.46 | 3.45±0.60 | 0.10±0.04 | 0.06±0.03 | 3.73±0.49 | 3.69±0.60 |
| **5** | -- | -- | -- | -- | -- | -- | -- | -- |
| **6** | 2.10±0.23 | 0.45±0.08 | 5.31±0.26 | 7.33±0.46 | 0.08±0.01 | 0.10±0.02 | 7.49±0.34 | 7.88±0.47 |
| **7** | 0.63±0.19 | 0.23±0.04 | 3.80±0.19 | 4.47±0.32 | 0.07±0.01 | 0.03±0.01 | 4.50±0.27 | 4.73±0.33 |
| **8** | 0.86±0.16 | 0.26±0.07 | 2.84±0.21 | 3.42±0.35 | 0.17±0.03 | 0.06±0.01 | 3.87±0.27 | 3.74±0.36 |
| **9** | 0.85±0.17 | 0.53±0.14 | 4.27±0.33 | 5.03±0.51 | 0.11±0.02 | 0.09±0.02 | 5.24±0.37 | 5.66±0.53 |
| **10** | 0.53±0.12 | 0.44±0.15 | 4.22±0.36 | 5.22±0.69 | 0.12±0.03 | 0.07±0.02 | 4.87±0.38 | 5.73±0.71 |
| **11** | 2.21±1.27 | 1.61±1.04 | 6.64±1.56 | 7.29±1.91 | 0.37±0.18 | 0.35±0.16 | 9.22±2.02 | 9.25±2.18 |
| 12 | 0.44±0.07 | 0.28±0.07 | 3.89±0.28 | 4.54±0.50 | 0.11±0.02 | 0.11±0.04 | 4.44±0.29 | 4.93±0.51 |
| 13 | -- | -- | -- | -- | -- | -- | -- | -- |
| 14 | 4.01±1.61 | 4.46±2.21 | 13.51±2.13 | 15.53±2.85 | 1.46±0.57 | 1.19±0.57 | 18.98±2.73 | 21.18±3.65 |
| 15 | 2.49±0.73 | 1.16±0.55 | 5.11±0.59 | 5.90±0.84 | 0.85±0.26 | 0.71±0.61 | 8.44±0.98 | 7.77±1.17 |
| 16 | 2.33±0.59 | 1.11±0.38 | 6.89±0.69 | 8.46±1.10 | 0.72±0.18 | 0.43±0.13 | 9.95±0.93 | 9.99±1.18 |
| 17 | 8.63±2.06 | 3.36±1.19 | 10.25±0.94 | 12.73±1.51 | 2.27±0.59 | 1.22±1.00 | 21.14±2.34 | 17.31±2.17 |
| 18 | 1.55±0.81 | 1.07±0.68 | 6.02±1.38 | 6.61±1.69 | 0.34±0.17 | 0.27±0.14 | 7.91±1.61 | 7.94±1.83 |
| 19 | 2.07±0.49 | 1.08±0.24 | 6.70±0.43 | 8.12±0.64 | 0.32±0.05 | 0.67±0.12 | 9.10±0.66 | 9.87±0.69 |
| 20 | 1.92±0.55 | 1.17±0.44 | 7.62±0.85 | 8.38±1.15 | 0.34±0.09 | 0.28±0.09 | 9.87±1.02 | 9.83±1.24 |
| 21 | 0.48±0.07 | 0.15±0.03 | 2.63±0.17 | 3.41±0.29 | 0.13±0.05 | 0.23±0.04 | 3.25±0.19 | 3.79±0.30 |

## 4. Summary and Conclusions

This study presents a detailed analysis of 14 high-resolution shocks in the distant outer heliosphere observed by the SWAP instrument onboard *New Horizons*. These shocks are observed over a heliocentric distance range of ~52-60 au. The distant interplanetary shocks observed over this distance are relatively weak, with PUI density compression ranging from ~1.2 – 1.9. The weakest shock was S11, observed on 2022 Dec 28 with a shock speed of ~463 km s$^{-1}$, and the strongest shock was S21, with a shock speed of ~400 km s$^{-1}$. The upstream sonic Mach number ranges from 0.69 (for shock S12) to 2.00 (for shock S17). Around 64% of them (9 shocks) have an upstream sonic Mach number greater than one. The sonic Mach number for all the shocks decreases from upstream to downstream. In addition, seven shocks with supersonic upstream flow transitioned to subsonic flow downstream, and the remaining two have downstream sonic Mach numbers less than one, considering their 1-sigma uncertainty. In general, most of these shocks are relatively narrower compared to the earlier six high-resolution fast-forward shocks analyzed by McComas et al. (2022). The shock transition scale (shock width) ranges from ~0.003 to 0.14 au, which is much larger than the upstream advective gyroradius of PUIs (~12 to 550 $r_g$). The compression of SW density from upstream to downstream in the newer shocks is somewhat clear compared to the earlier six high-resolution shocks. However, the SW ions are still not compressed by the same amount as PUIs compression.

We also presented a statistical analysis of shock parameters in the distant outer heliosphere using 19 fast-forward shocks observed between ~49.5-60 au. This also includes six earlier high-resolution shocks analyzed by McComas et al. (2022). Most distant interplanetary shocks in this distance range have a compression ratio of ~1.3 and a shock speed of ~60 km s$^{-1}$ in the upstream plasma frame. The compression in SW density appears to be slowly increasing with the PUI density compression, with a weaker correlation between them. The shock width is (i) independent of the shock compression ratio, (ii) strongly anti-correlated with the shock speed in the upstream plasma frame, and (iii) weakly anti-correlated with the upstream sonic Mach number. The energy flux of SW ions is decreased across the shock while the energy flux of PUIs is increased. In addition, the change in energy flux across the shock increases with the shock compression ratio and the shock speed in the upstream plasma frame. Furthermore, the change in the energy flux increases with the upstream plasma flow speed in the shock frame.

SWAP will continue to make high-resolution observations of PUI-mediated distant interplanetary shocks as *New Horizons* continues its journey toward the HTS. When *New Horizons* crosses the HTS, SWAP's PUI measurements will be fundamental in understanding the structure and physical processes of the HTS, including energy partitioning at the HTS. Moreover, SWAP's PUI measurements in the heliosheath will enable us to understand how the PUI distribution evolves in the heliosheath and their role in the production of heliospheric ENAs, which are currently being measured by IBEX and will soon be measured by IMAP.

*Acknowledgments.* We acknowledge the support of the *New Horizons* mission (M99023MJM; PU-AWD1006357) as part of NASA's New Frontier program and the IMAP mission as part of NASA's Solar Terrestrial Program (STP) mission line (80GSFC19C0027). The *New Horizons* SWAP data is available at https://spacephysics.princeton.edu/missions-instruments/swap/pui-data-2025, and the magnetic field data at one au is available at https://omniweb.gsfc.nasa.gov/. B.S. thanks Laxman Adhikari for the helpful discussions.

## References

Baliukin, I. I., Izmodenov, V. V., & Alexashov, D. B. 2022, MNRAS, 509, 5437

Bzowski, M., Möbius, E., Tarnopolski, S., Izmodenov, V., & Gloeckler, G. 2009, SSRv, 143, 177
Chen, J. H., Bochsler, P., Möbius, E., & Gloeckler, G. 2014, JGR, 119, 7142
Decker, R. B., Krimigis, S. M., Roelof, E. C., et al. 2008, Natur, 454, 67
Dialynas, K., Galli, A., Dayeh, M. A., et al. 2020, ApJL, 905, L24
Elliott, H. A., McComas, D. J., Zirnstein, E. J., et al. 2019, ApJ, 885, 156
Gkioulidou, M., Opher, M., Kornbleuth, M., et al. 2022, ApJL, 931, L21
Kornbleuth, M., Opher, M., Zank, G. P., et al. 2023, ApJL, 944, L47
Kumar, R., Zirnstein, E. J., & Spitkovsky, A. 2018, ApJ, 860, 156
Lee, M. A., Shapiro, V. D., & Sagdeev, R. Z. 1996, JGR, 101, 4777
Livadiotis, G. 2015, ApJ, 809, 111
Livadiotis, G., McComas, D. J., & Shrestha, B. L. 2024, ApJ, 968, 66
McComas, D., Allegrini, F., Bagenal, F., et al. 2008, SSRv, 140, 261
McComas, D. J., Allegrini, F., Bochsler, P., et al. 2009, Sci, 326, 959
McComas, D. J., Christian, E. R., Schwadron, N. A., et al. 2018, SSRv, 214, 116
McComas, D. J., Elliott, H. A., & Schwadron, N. A. 2010, JGR, 115, A03102
McComas, D. J., Shrestha, B. L., Livadiotis, G., et al. 2025, ApJ, 980, 154
McComas, D. J., Shrestha, B. L., Swaczyna, P., et al. 2022, ApJ, 934, 147
McComas, D. J., Swaczyna, P., Szalay, J. R., et al. 2021, ApJS, 254, 19
McComas, D. J., Zirnstein, E. J., Bzowski, M., et al. 2017, ApJS, 233, 8
Mostafavi, P., Zank, G. P., & Webb, G. M. 2018, ApJ, 868, 120
Neugebauer, M. 2013, SSRv, 176, 125
Randol, B. M., Elliott, H. A., Gosling, J. T., McComas, D. J., & Schwadron, N. A. 2012, ApJ, 755, 75
Randol, B. M., McComas, D. J., & Schwadron, N. A. 2013, ApJ, 768, 120
Richardson, I. G. 2004, SSRv, 111, 267
Richardson, I. G. 2018, LRSP, 15, 1
Richardson, J. D., Kasper, J. C., Wang, C., Belcher, J. W., & Lazarus, A. J. 2008a, Natur, 454, 63
Richardson, J. D., Liu, Y., & Wang, C. 2008b, AdSpR, 41, 237
Rucinski, D., & Bzowski, M. 1995, A&A, 296, 248
Shrestha, B. L., Zirnstein, E. J., & Heerikhuisen, J. 2020, ApJ, 894, 102
Shrestha, B. L., Zirnstein, E. J., Heerikhuisen, J., & Zank, G. P. 2021, ApJS, 254, 32
Shrestha, B. L., Zirnstein, E. J., & McComas, D. J. 2023, ApJ, 943, 34
Shrestha, B. L., Zirnstein, E. J., McComas, D. J., et al. 2024, ApJ, 960, 35
Sokół, J. M., Kubiak, M. A., & Bzowski, M. 2019, ApJ, 879, 24
Swaczyna, P., McComas, D. J., Zirnstein, E. J., et al. 2020, ApJ, 903, 48
Treumann, R. A. 2009, ARA&A, 17 (Springer), 409
Vasyliunas, V. M., & Siscoe, G. L. 1976, JGR, 81, 1247
Wang, C., & Richardson, J. D. 2002, GRL, 29, 1181
Zank, G. P., Heerikhuisen, J., Pogorelov, N. V., Burrows, R., & McComas, D. 2010, ApJ, 708, 1092
Zank, G. P., Pauls, H. L., Cairns, I. H., & Webb, G. M. 1996, JGR, 101, 457
Zirnstein, E. J., Kumar, R., Bandyopadhyay, R., et al. 2021, ApJL, 916, L21
Zirnstein, E. J., McComas, D. J., Kumar, R., et al. 2018, PhRvL, 121, 075102
Zirnstein, E. J., Möbius, E., Zhang, M., et al. 2022, SSRv, 218, 28

Zirnstein, E., Kumar, R., Swaczyna, P., et al. 2023 (Research Square)